\documentclass[fleqn,usenatbib]{mnras}

\usepackage{newtxtext,newtxmath}

\usepackage[T1]{fontenc}

\DeclareRobustCommand{\VAN}[3]{#2}
\let\VANthebibliography\thebibliography
\def\thebibliography{\DeclareRobustCommand{\VAN}[3]{##3}\VANthebibliography}

\usepackage{graphicx}	
\usepackage{amsmath}	

\def\kpc{\,\mathrm{kpc}}

\def\kms{\,\mathrm{km\,s}^{-1}}

\newcommand{\vect}[1]{\boldsymbol{#1}}

\def\vlos{{v_\parallel}}

\def\AllWISE{{\it AllWISE}}
\def\Gaia{{\it Gaia}}

\title[Tangled warp II: The red clump]{The tangled warp of the Milky Way II: The red clump}

\author[Turner \& McMillan]{
	James E. Turner, 
Paul J. McMillan
\thanks{E-mail: paul.mcmillan@le.ac.uk}
\\
School of Physics \& Astronomy, University of Leicester, University Road, Leicester LE1 7RH, UK\\
}

\date{Accepted XXX. Received YYY; in original form ZZZ}

\pubyear{\the\year{}}

\begin{document}
\label{firstpage}
\pagerange{\pageref{firstpage}--\pageref{lastpage}}
\maketitle

\begin{abstract}
    The Milky Way's disc is warped in a way that varies depending on the age of the population considered. We, for the first time, trace that kinematics of the warp in red clumps stars. This means that our sample is drawn from a consistent population across the disc, composed of standard candles that allow for precise distance estimates but, unlike samples of Cepheid variables, is relatively old. We find that, consistent with other populations, the red-clump warp is precessing at ${\sim}12 \kms\kpc^{-1}$, comparable to the angular velocity of stars around the Galaxy in the outer reaches of the disc. Unlike younger populations, the kinematic signature is offset from that which we would expect given the stellar positions.
    This reinforces the conclusion that the Milky Way's warp is dynamically evolving, making this a necessary property of any model that seeks to explain its creation.
\end{abstract}

\begin{keywords}
Galaxy: disc -- Galaxy: kinematics and dynamics -- Galaxy: structure
\end{keywords}



\section{Introduction}

\noindent The Milky Way's disc is, we know, far from featureless. In addition to the bar and spiral arms, it has long been known to be substantially warped, with the first proof of this coming from H{\textsc i} mapping \citep{kerr_magellanic_1957}. In the era of \Gaia\ data \citep{gaia_collaboration_mission_2016}, we are increasingly able  to map its geometry and motion with stellar tracers \citep[e.g.,][]{poggio_galactic_2018, dehnen_a_twisted_2023, uppal_clump_2024}.

Observations of external galaxies have shown that warping is common, occurring in 40-50\% of edge-on spiral galaxies \citep{richter_asymmetries_1994, reshetnikov_statistics_1998}. The presence of Milky Way-like warps in these systems implies that our Galaxy is not atypical, but rather that warping is a natural outcome of disc galaxy evolution. While the origin of warping within our Galactic disc is still an open question, current leading theories include a misaligned or rotating dark matter halo \citep{han_tilted_2023}, accretion of gas with angular momentum not aligned with the Galactic disc \citep{jiang_infall_1999}, or tidal interactions between the Milky Way and companion galaxies \citep{khanna_dr2_2019, laporte_footprints_2019}. 

Since we can view the Milk Way's warp in far more detail than any other, it provides a great test-bed to understand the formation and evolution of warps in galactic discs. The use of different tracers has shown us that the signature of the Milky Way's can be found for a wide range of stellar ages \citep[e.g.,][]{poggio_galactic_2018} but that its behaviour depends significantly on the age of the tracers used \citep[e.g.,][]{amores_evolution_2017,uppal_clump_2024}. The warp appears to be twisted, which is usually parametrized by the `line of nodes'' (LON). The LON represents the line in which the stellar disc of the Galaxy intersects the flat midplane, marking the transition where the disc crosses upward or downward at the ascending and descending nodes, respectively. The position of the LON has been shown to vary with radius using the positions and velocities of the young Cepheid population \citep[e.g.,][]{dehnen_a_twisted_2023}, and the velocities of stars in the bulk population, which contains stars of all ages \citep[][henceforth Paper I]{jonsson_twisted_2024}. In the case of Paper I, the sample was all stars with measured line-of-sight velocities in \Gaia\ Data Release 3 \citep[DR3,][]{ gaia_collaboration_summary_2023, katz_gaia_2023}. However, no-one has yet found a significant twist in the warp as seen in the positions of older stars, perhaps because of the dust extinction which hampers the analysis \citep{uppal_clump_2024, khanna_unlimited_2025}.


Different tracers not only represent different ages and/or metallicities of stars, they also offer different advantages for studies. Classical Cepheids or Upper Main Sequence (UMS) stars are young and inherit the motion of the molecular gas they have formed from, largely unaffected by later dynamical perturbations. Because of this, they trace the young stellar disc and spiral arm architecture \citep[e.g.][]{skowron_mapping_2019, poggio_greatwave_2025}. It is common to study the Cepheids population particularly \citep[e.g.][]{chen_intuitive_2019, dehnen_a_twisted_2023, cabrera_cepheids_2024}, because the period-luminosity relation they follow makes them excellent standard candles \citep{leavitt_cepheid_1912}, and therefore we are able to accurately and reliably determine their distances. 

To understand stars with a broader range of ages, which will have orbits that are significantly different from those they had at birth, there is a similar choice to be made. One can look at the full observable population, or a very broad subset such as the red giant branch \citep[e.g.,][Paper I]{amores_evolution_2017, poggio_galactic_2018}. Alternatively, one can focus purely on red clump (RC) stars, which are good standard candles and have a broad spread of ages \citep[e.g.][]{Girardi_RC_2016}. RC stars are abundant across the Galaxy and therefore provide both precise distance estimates and a high density of data points, greatly increasing our ability to map the large-scale kinematics of the Milky Way \citep[e.g.][henceforth K25]{williams_wobbly_2013, uppal_clump_2024, khanna_unlimited_2025}. However, galactic extinction makes it very difficult to determine the properties of the RC warp from the positions of stars alone. Despite careful measurement of the selection function applying to their sample, K25 were unable to determine the LON of their sample, and had to fix it near the Galactic anticentre (Galactocentric $\phi=180^\circ$), very close to the position of the LON found by \citet{uppal_clump_2024}.

\begin{figure*}
	\includegraphics[width=0.67\columnwidth]{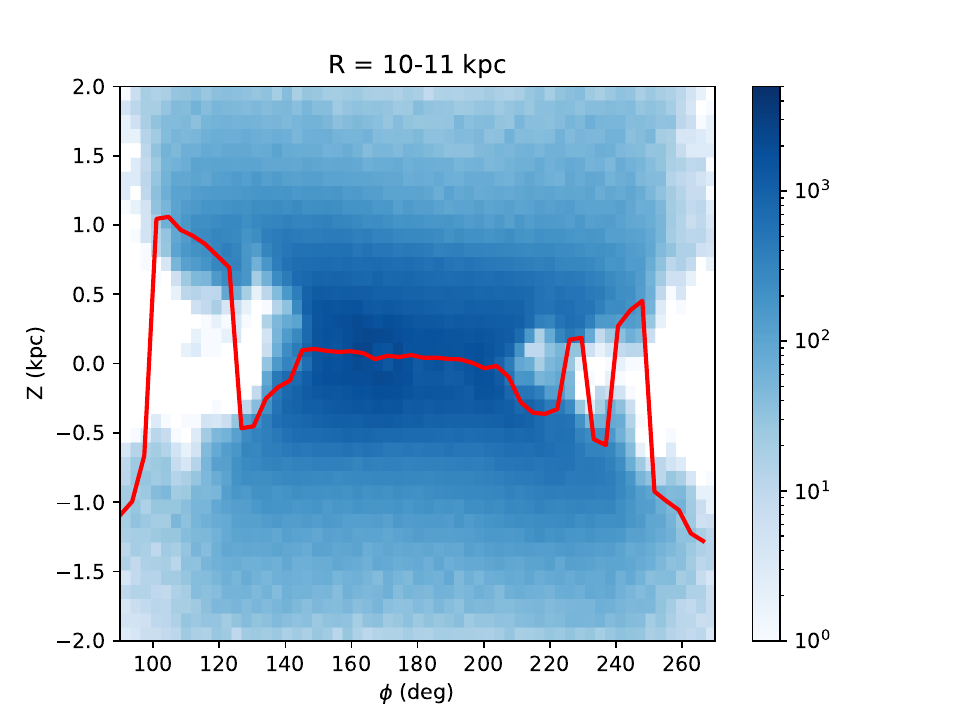}
	\includegraphics[width=0.67\columnwidth]{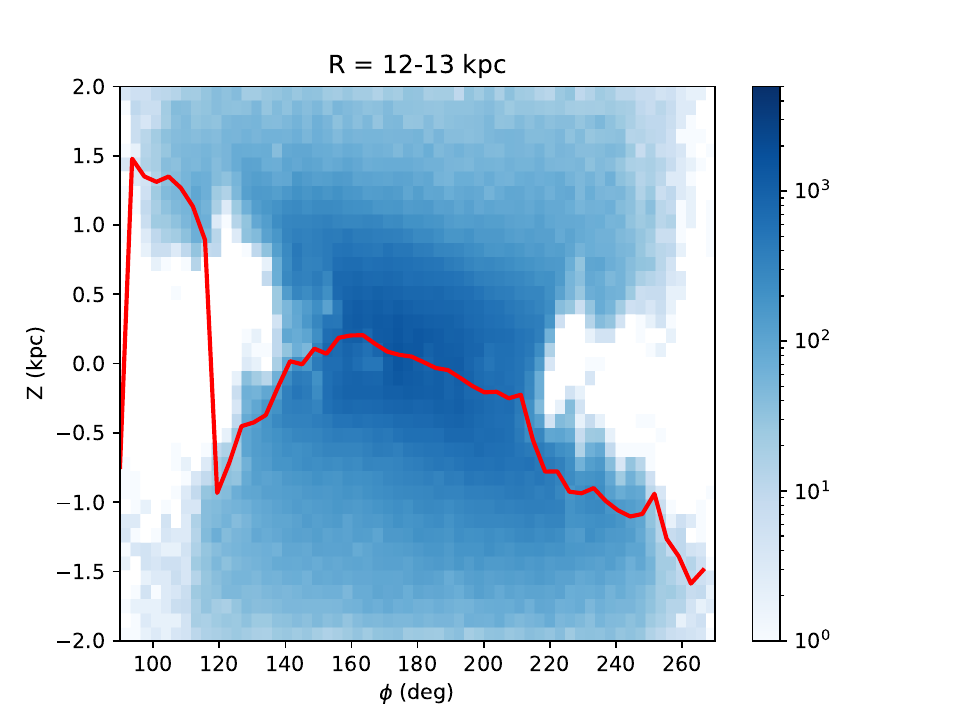}
	\includegraphics[width=0.67\columnwidth]{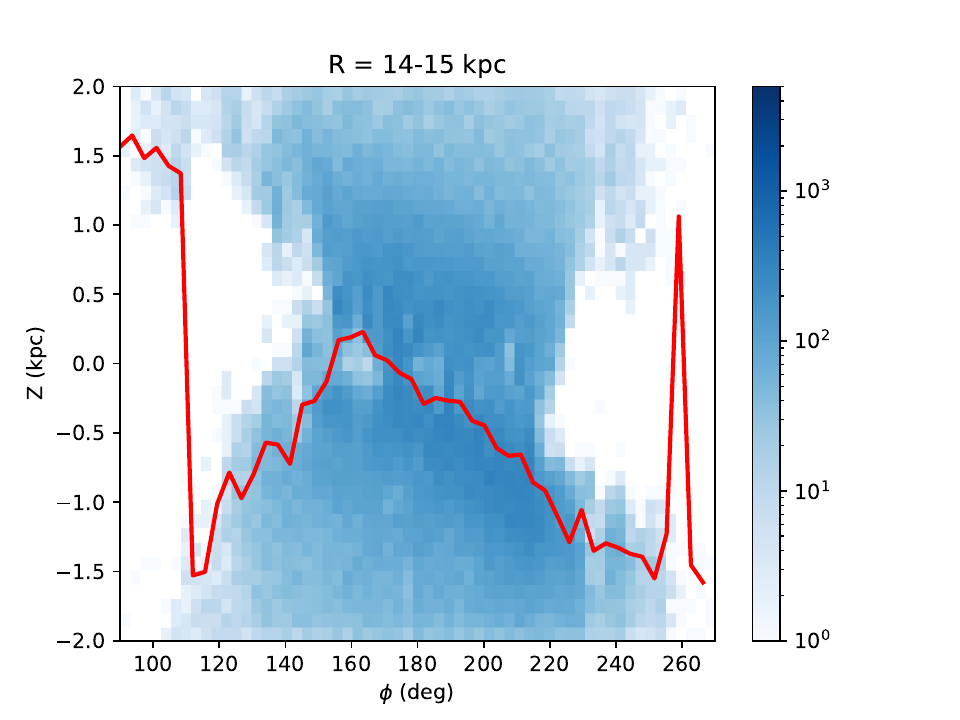}
    \caption{Vertical distribution of RC stars within the dataset with respect to Galactocentric azimuth. Panels from left to right show radial slices at $R=10-11$ kpc, $12-13$ kpc, and $14-15$ kpc. The red line indicates the median stellar vertical position ($Z$), with the colour intensity representing the stellar number density. Distinct fluctuations of the median height of the stellar sample are explained by artificial skewing arising from high dust extinction obscuring the view to stars along specific lines of sight. The grid resolution is $3^\circ$ in azimuth and 100 pc in $Z$.}
    \label{fig:zphiplot}
\end{figure*}

In a simple model, either simple tilted rings or Jeans modelling of the population, we expect that the average vertical velocities of stars are related to the geometric shape of the warp. The LON should also be the position of the greatest vertical velocity. For the Cepheid population this appears to be close to true \citep{dehnen_a_twisted_2023, cabrera_cepheids_2024}, while in Paper I we found this was not the case for the bulk population, with the LON determined from the velocity data being close to $\phi = 135^\circ$, rather than the canonical $180^\circ$ for the geometric warp (i.e., offset by around $45^\circ$ in the direction of Galactic rotation).

In this study we investigate the kinematics of the Milky Way's warp as traced by red clump stars. This allows us to use more precise distance estimates and a more clearly defined sample of stars than the pot-pourri of stars in the full \Gaia\ DR3 line-of-sight velocity catalogue used in Paper I. This study therefore serves as an extension to the methodology established in Paper I. Here, we apply similar rigid, precessing disc models testing both a fixed LON and precession rate, as well as a ``twisted'' configuration where both parameters vary linearly with Galactocentric radius.


We structure our paper as follows: Section~\ref{sec:DataModel} describes the data we use and our modelling framework; Section~\ref{sec:RCwarp} describes the properties of the red clump warp found as a result of our modelling efforts; 
and Sections~\ref{sec:discussion}~\&~\ref{sec:conclusions} discuss our results and draw conclusions, respectively.

\section{Data and modelling} \label{sec:DataModel}
Our main sample is taken from K25, who made a photo-astrometric selection of red clump stars using \AllWISE\ photometry \citep{wright_wise_2010,cutri_allwise_2014} photometry and \Gaia\ DR3 photometry and astrometry \citep{riello_gaia_2021, lindegren_astrometric_2021}. This allowed them to create a sample of stars which can be assumed with high confidence to truly be from the red clump, and determine the selection function that describes how likely a red clump star in any region of the galaxy is to enter the catalogue. To ensure the high quality of the RC sample, they also applied several masks and quality cuts to the data. First, all sources within a Galactocentric radius of $R<3$ kpc were removed to avoid the complex kinematics and contamination of the Galactic bulge and central bar. Second, sources beyond a vertical distance of 2 kpc from the Galactic midplane ($|Z|>2$ kpc) were removed to exclude contamination from high-latitude sources in the Large and Small Magellanic Clouds. Third, the worst 0.5\% of the data was excluded by requiring the photometric uncertainty in the AllWISE $W_1$ band to be $\sigma_{W1}<0.1$. Finally, a faint-end magnitude limit of $\rm G=17.5$ was imposed. These cuts generated a high-quality sample of $8\, 729\, 562$ RC stars spatially centred on the Sun's position within the Galaxy.

From the data K25 were unable to determine the position of the LON, but they did measure the height of the red-clump warp as a function of radius, assuming a position for the LON near $180^\circ$ in the Galactic disc. The difficulty in determining the position of the LON is well illustrated by Figure~\ref{fig:zphiplot} which shows the density of RC stars in bins of $Z$ and $\phi$ at different radii. It is obvious from this basic view that dust extinction causes large holes in the data that vary substantially in $\phi$, and are therefore very difficult to account for, even with a well estimated selection function.

In this work we therefore focus entirely on the kinematics of the sample, particularly on the vertical velocity $v_Z$. This is naturally much less susceptible to the effects of dust, as we only require that the samples' average $v_Z$ is an unbiassed estimate of the true average $v_Z$ function of $R$ and $\phi$, integrated over all $Z$. We explore likely breaches of this assumption in Appendix~\ref{sec:checks}.


\subsection{Missing line-of-sight velocities} \label{sec:MissLOS}
Out of the sample of $8\,729\,562$ red clump stars, only $4\,535\,582$ have measured line-of-sight velocities ($\vlos$) in \Gaia\ DR3. We are therefore forced to either work with a subset of the stars that covers a smaller area of the disc, or make assumptions about the kinematics of the disc. Here we follow the approach taken in \cite{mcmillan_disturbed_2022}, in which the average Galactocentric radial velocity is assumed to be $\overline{v}_R\approx 0$ throughout the disc. 

To derive these missing line-of-sight velocities, we take the Sun's Galactocentric radius as $R_{\odot}=8.178$ kpc \citep{2018A&A...615L..15G} and its height above the midplane as $Z_{\odot}=20$ pc \citep{humphreys_suns_1995}, along with velocities $\overline{\vect{v}}_{\odot}=(v_{R,\odot},v_{\phi,\odot},v_{Z,\odot})=(11.1,241.2,7.25)\kms$ \citep{schonrich_local_2010,eilers_circular_2019}. In Cartesian coordinates, the true Galactocentric velocity of a star, $\vect{v}_{GC}$, is the vector sum of its motion derived from its proper motion and the motion derived from its line-of-sight velocity. Letting $\vect{v}_{GC,0}$ be the Galactocentric velocity vector under the assumption that $\vlos=0$, the true Galactocentric velocity can be expressed as a linear combination of this baseline velocity and the contribution from the true line-of-sight velocity, given by:

\begin{flalign}
& \vect{v}_{GC}=\vect{v}_{GC,0}+\vlos\hat{\vect{n}} & 
\label{eq:LOS_v_GC}
\end{flalign}

\noindent where $\hat{\vect{n}}$ 
is the unit vector pointing from the Sun to the star. In the Galactocentric cylindrical frame, the planar radius vector is $\vect{R}=(x,y,0)$, and the associated unit vector in the radial direction is $\hat{\vect{e}}_R=(x/R,y/R,0)$, for cylindrical radius $R$. Therefore, the Galactocentric radial velocity, $v_R$, is the projection of the total Galactocentric velocity onto $\hat{\vect{e}}_R$:

\begin{flalign}
& v_R=\vect{v}_{GC} \cdot \hat{\vect{e}}_R=v_{R,0}+\vlos B & 
\label{eq:LOS_v_R}
\end{flalign}

\noindent where $v_{R,0}=\vect{v}_{GC,0}\cdot\hat{\vect{e}}_R$ is the Galactocentric radial velocity contribution from proper motions and solar motion. The parameter $B$ is a geometric projection factor defined as
\begin{equation}
    B = \hat{\mathbf{n}} \cdot \hat{\vect{e}}_R
\end{equation}
the dot product of the line-of-sight unit vector and the Galactocentric radial unit vector. By imposing the assumption that stars, on average, have $v_R=0$, the equation can be rearranged to solve for the unknown line-of-sight velocity,

\begin{flalign}
& \vlos=-\frac{v_{R,0}}{B}. & 
\label{eq:LOS_v_los}
\end{flalign}

\begin{figure*}
	\includegraphics[width=2\columnwidth]{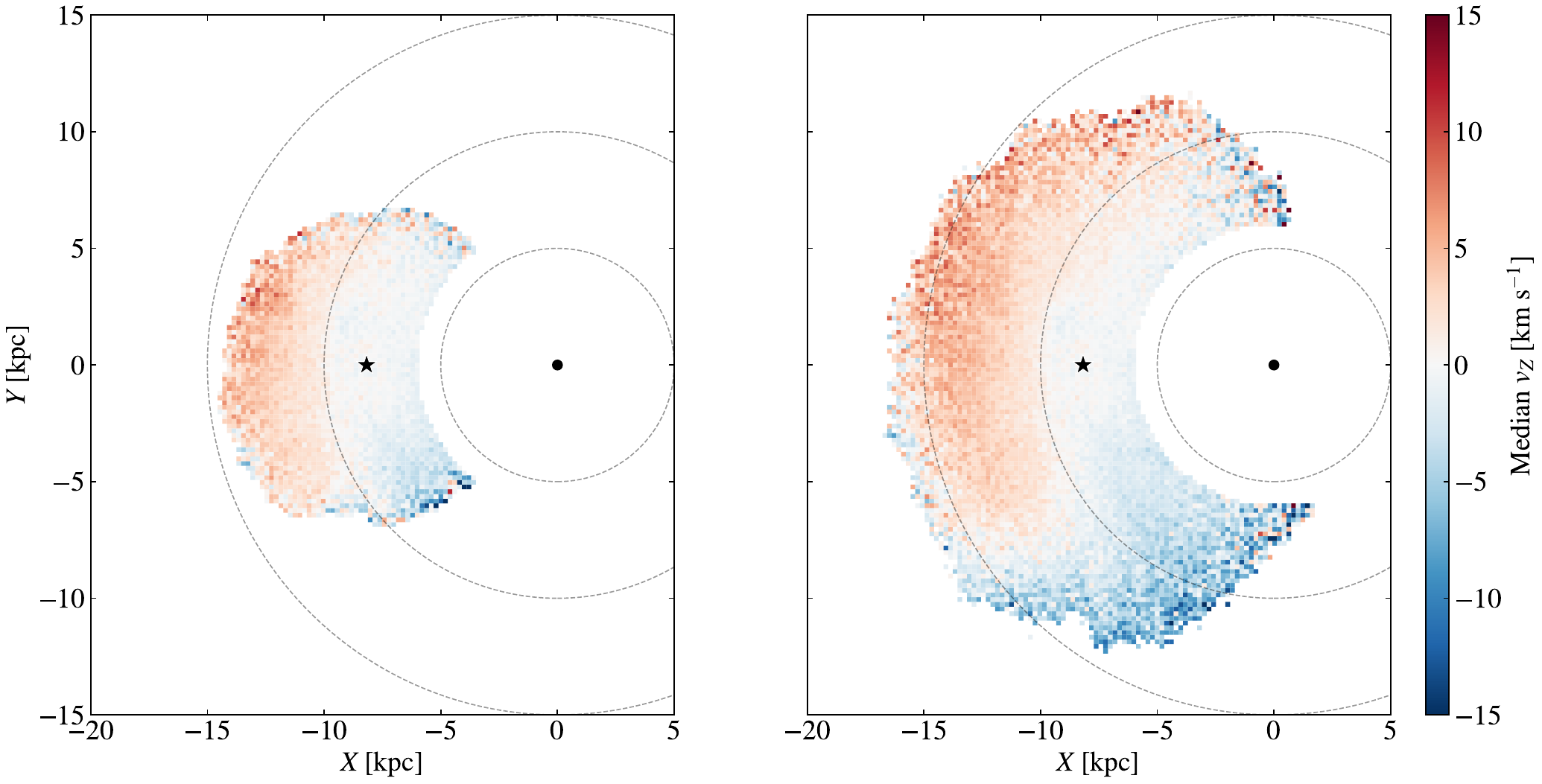}
    \caption{Top-down median vertical velocity distribution ($\overline{v}_Z$) for RC sample in Cartesian coordinates (the Sun and Galactic centre are indicated; Galactocentric circles differ by 5 kpc in radius). The left panel displays the spatial coverage of stars with complete 3D velocity data. The right panel shows the expanded spatial footprint when including estimated line-of-sight velocities from the assumption of zero Galactocentric radial velocity ($v_{R,gal}\approx0$). Bins contain a minimum of 50 stars and any bins with fewer than 50 stars were not plotted.}
    \label{fig:v_los_rangeincrease} 
\end{figure*}

\noindent The stability of the $\vlos$ estimate depends on the projection factor, $B$, leading to specific regions in which this assumption works exceptionally well. For example, when looking directly toward or away from the Galactic centre ($l=0^\circ,180^\circ$), the Galactocentric radius aligns with the line of sight, allowing the projection factor to reach its maximum absolute value ($|B|\approx1$). In contrast, when the line of sight is perpendicular to the Galactocentric radial vector the projection factor approaches zero, for example, close to the Sun this is the direction towards or away from Galactic rotation ($l=90^\circ,270^\circ$). To mitigate the division by values close to zero, which results in line-of-sight velocities approaching infinity, a geometric filter was applied, and estimations were only accepted for stars where the projection factor satisfied $|B|>0.1$.

This approximation works relatively well for vertical velocities in the disc, since outside the solar neighbourhood the line-of-sight velocity almost always provides a small fraction of the vertical velocity. However, the line-of-sight velocity is often a major component of $v_\phi$ for disc stars. To ensure we have a reasonable estimate of $\overline{v}_\phi$ for any region of the disc we sample, we take advantage of the fact that we expect it to have a relatively small variation with $\phi$. For a given radius, we take the $\overline{v}_\phi$ in all bins to be the median $v_\phi$ in the range $160^\circ<\phi<200^\circ$, with these limits chosen so that the line-of-sight component is expected to provide a small fraction of $v_\phi$ for any star \citep[e.g.,][]{mcmillan_disturbed_2022}.

Furthermore, it must be noted that the calculation breaks down in regions where the assumption of zero Galactocentric radial velocity is violated. Toward the Galactic centre, the presence of the Galactic bar heavily perturbs nearby orbits, inducing high eccentricities. Similarly, at high Galactic latitudes, this planar model projects 3D sight lines onto a 2D kinematic constraint. For stars at large distances away from the midplane ($|b|\gg0^\circ$), the sight line introduces a steep angle, causing $\hat{n}$ to become increasingly orthogonal to the planar vector $\hat{e}_R$ and resulting in an overestimation of the line-of-sight velocity. A discussion of the errors introduced when this assumption is incorrect is presented in Appendix~\ref{sec:checks}.

Even with these limitations, this assumption of zero Galactocentric radial velocity provides sufficiently accurate line-of-sight velocity estimates for 3 860 894 sources, increasing the 6D phase-space coordinate coverage of the sample to 96\%. The resulting increase in spatial coverage is illustrated in Figure \ref{fig:v_los_rangeincrease}. We explore the robustness of this technique to realistic breaches of our assumptions in Sec~\ref{sec:checks}.

\subsection{Modelling} \label{ssec:Model}
Our model is very similar to the ``Classical'' models in Paper I, with some minor adjustments. This makes them akin to modelling efforts that have been used on numerous occasions \citep[e.g.,][]{drimmel_galactic_2000, poggio_evidence_2020}. We assume the midplane of the Milky Way's disc can be approximated as a series of coupled tilted rings, precessing with angular velocity $\omega_p$. The vertical position of the midplane in Galactocentric cylindrical coordinates as a function of parameters $R$ and $\phi$ is therefore: 

\begin{flalign}
& Z_0(R,\phi)=h(R)\sin(\phi-\phi_w(R)-\omega_p(R)t), & 
\label{warpshape}
\end{flalign}

\noindent where $t$ is time, $h(R)$ is the largest displacement of the warp at a given Galactocentric radius $R$, $\phi$ is the Galactocentric azimuth, and $\phi_w(R)$ is the LON of the warp. Given the direction of Galactic rotation, $\phi_w(R)$ is on the far side of the Galaxy, so we refer instead to $\phi^{'}_w(R) = \phi_w(R) +180^\circ$

To model the variation in the warp height, we employ a piecewise-linear model rather than a single continuous equation to describe the entire disc; splitting the radii into 1 kpc intervals from $R = 7$ to 16 kpc (denoted as parameters $h_7,h_8,...,h_{16}$). This approach affords us higher flexibility in the model and avoids forcing a specific assumed shape. 

Two specific configurations of this warp shape are tested. The first is the fixed model, where both the LON ($\phi_w^{'}$) and the precession rate ($\omega_p$) stay constant across the entire disc. The second configuration allowed both of these parameters to vary linearly with Galactocentric radius and is denoted as the twisted model, accommodating a ``twisted'' warp where the LON and precession rate are defined as: 

\begin{flalign}
& \phi_w^{'}(R)=\phi_{w,0}^{'} + \phi_{w,1}(R-R_s)& 
\label{twistedlon}
\end{flalign}

\begin{flalign}
& \omega_p(R)=\omega_{p,0} + \omega_{p,1}(R-R_s),& 
\label{twistedprecession}
\end{flalign}

\noindent where $R_s$ is the reference radius, which is placed at 10 kpc as a representative midpoint for our outer-disc data. Both $\phi_{w,0}^{'}$ and $\omega_{p,0}$ are therefore the base LON and precession rate at this reference radius, with $\phi_{w,1}$ and $\omega_{p,1}$ representing their gradient parameters, describing how they vary with respect to Galactocentric radius. To improve upon the models seen in Paper I and increase the physical accuracy of the twisted model, we place constraints on where and how much certain parameters can vary. Firstly, because no significant warping is observed in the inner disc (R\,$<$\,10 kpc), we constrain the LON and precession rate to vary only at radii beyond 10 kpc. Secondly, we constrain the precession gradient to be within $\pm$ 2 km\,s$^{-1}\, \rm kpc^{-2}$. Finally, we force the warp height to increase monotonically with Galactocentric radius. These constraints on the twisted model prevent the Markov Chain Monte Carlo algorithm from exploring mathematically correct yet unphysical solutions. 

\subsection{Expected Vertical Velocities}
\noindent To calculate the expected vertical velocities for these models, we use Jeans' first equation \citep{jeans_theory_1915} to determine the average velocity of the collisionless system under the assumption that the warp shape is fixed but precessing. The equation in cylindrical polar coordinates can be written as:

\begin{flalign}
& \frac{\partial n}{\partial t}+\frac{\partial (n\overline{v}_R)}{\partial R}+\frac{1}{R}\frac{\partial (n\overline{v}_\phi)}{\partial \phi}+\frac{\partial (n\overline{v}_Z)}{\partial Z}=0,& 
\label{jeansequation}
\end{flalign}

\noindent where $n$ is the stellar number density.

We can neglect the variation of $\overline{v}_\phi$ with azimuth, and of $\overline{v}_Z$ with height above or below the midplane ($\partial\, \overline{v}_Z/\partial Z=0$). Paper I also showed that attempting to include the term in $\overline{v}_R$ leads to unrealistic and poorly-fitting results. We are left with

\begin{equation}
\begin{split}
\overline{v}_{Z,\rm mod} =\;& 
\left( \frac{\overline{v}_\phi}{R} - \omega_p(R) \right) h(R) \\
&\times 
\cos\!\left(\phi-\phi_w^{'}(R)-\omega_p(R)t\right)
\end{split}
\label{m1velocity}
\end{equation}

\noindent Equation \ref{m1velocity} describes the mean expected vertical velocities for the fixed and twisted models we use in this study to fit the observed and estimated data. It must also be noted that the LON found by the fitting algorithm generally favoured the warp on the opposing side of the disc; therefore, we add $180^\circ$\ to the calculated LON for all models to represent the symmetric side of the warp that is closer to the Sun.

\subsection{Fitting the Data}

\noindent To fit the data to the described models, we have to determine a posterior probability density function (PDF) for our model parameters. To approximate this, we employ the Markov chain Monte Carlo (MCMC) algorithm created by \citet{goodman_ensample} and conveniently packaged as \texttt{emcee} by \citet{foreman-mackey_emcee_2013}.
We assume flat priors for all parameters within physically allowed regions, and determine a log-likelihood based on the binned data, with bins spanning $0.5\kpc$ in $R$ and $10^\circ$ in $\phi$.

The total uncertainty for each spatial bin, $\sigma_{tot}$, was calculated by adding the statistical uncertainty, determined through bootstrapping, ($\sigma_{stat}$), a constant systematic uncertainty ($\sigma_{sys}$) of $0.7 \kms$, and the propagated model uncertainty ($\sigma_{mod}$, found from the statistical uncertainty in $\overline{v}_\phi$) in quadrature: 

\begin{flalign}
& \sigma_{\rm tot} = \sqrt{\sigma_{\rm stat}^2 + \sigma_{\rm mod}^2 + \sigma_{\rm sys}^2}. &
\label{totaluncertainty}
\end{flalign}

\noindent The log-probability sampled by the walkers is defined as:

 \begin{figure*}




    \includegraphics[width=2\columnwidth]{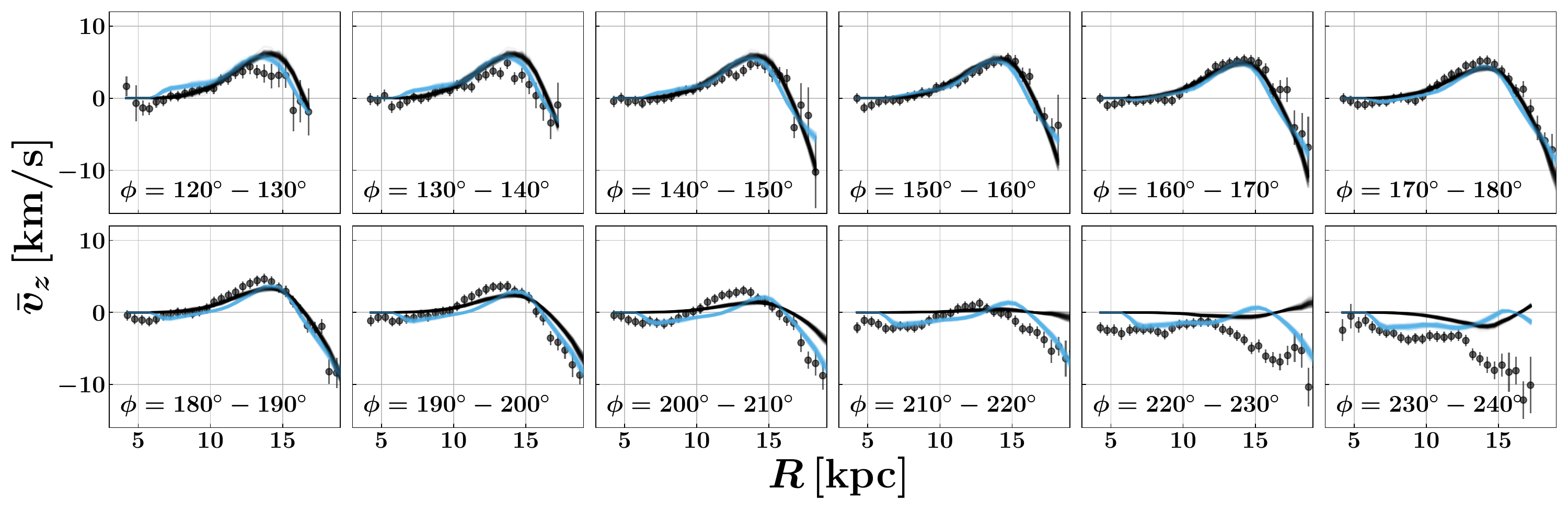}

    \caption{Fit of the expected vertical velocities ($\overline{v}_Z$) of the  twisted model (blue lines) and fixed model (black lines) to the dataset (black dots) as a function of Galactocentric radius ($R$) analysed in bins of 10 degrees Galactocentric azimuth ($\phi$). Observed data contains errors derived with bootstrapping. The models are represented by 100 randomly drawn curves from the final MCMC chains.}
    \label{fig:ModelDataComp} 
\end{figure*} 

\begin{flalign}
& \ln p \propto -\frac{1}{2}\sum \frac{(\overline{v}_{Z,\rm mod}-\overline{v}_{Z,\rm data})^2}{\sigma_{\rm tot}^2}. &
\label{loglikelihood}
\end{flalign}

\noindent where the sum is calculated over all spatial bins included in the sample.

The total number of parameters modelled using \texttt{emcee} includes 12 for the fixed model: warp height in 1 kpc sections from radii 7 to 16 kpc ($h_7,h_8,...,h_{16}$), LON ($\phi_w^{'}$), and precession rate ($\omega_p$). The twisted model adds two more parameters to this: specifically, the LON and precession rate gradients ($\phi_{w,1}$ and $\omega_{p,1}$ respectively), bringing the total dimensionality to 14 parameters.

To find these best-fitting parameters, \texttt{emcee} was used with 50 walkers exploring this parameter space over $20\, 000$ steps, discarding the first 20\% of the steps to account for the ``burn-in'' phase.

\section{The red clump warp} \label{sec:RCwarp}

In Figure~\ref{fig:ModelDataComp} we show a comparison between the model and data values for both the fixed and twisted models. Both models do a reasonable job of reproducing the vertical velocity trends associated with a warping disc. They closely match the data between $\phi=140^\circ$ and $200^\circ$, though data from outside of this range are less well matched by the model. Both models have $\overline{v}_Z$ values that are higher than the observed values at $\phi\lesssim140^\circ$ and $\phi\gtrsim200$. This is consistent with the results in Paper I, particularly for $\phi\gtrsim210$ where both datasets have negative $\overline{v}_Z$ in the outer radii, which the models struggle to capture.

It is likely that a key shortcoming of our models is that the warp cannot be entirely modelled with first-order Fourier terms, which force a strict global antisymmetry. Studies such as \cite{levine_vertical_2006} and \citealt{cabrera_cepheids_2024} have found that the Milky Way's warp is lopsided and has different warp amplitudes on the opposite sides, and therefore forcing an antisymmetric shape can lead to the model attempting to average out the two warp amplitudes, causing discrepancies at extreme azimuths. It may also be the case that we are failing to capture the behaviour of a disc that has an additional wave or corrugation \citep[e.g.][]{xu_rings_2015, poggio_greatwave_2025}.

Outside of these limitations, in these regions of extreme azimuths the twisted model is generally a slightly better fit, as one would expect since it has two further parameters to vary. Where the fixed model flattens out near $\phi\approx210^\circ-230^\circ$, the twisted model, by allowing the LON and precession to vary, introduces a phase shift in the outer disc. This twist allows more accurate downward velocity reproduction and may help provide evidence that the warp twists with Galactocentric radius.

\begin{figure}
	\includegraphics[width=\columnwidth]{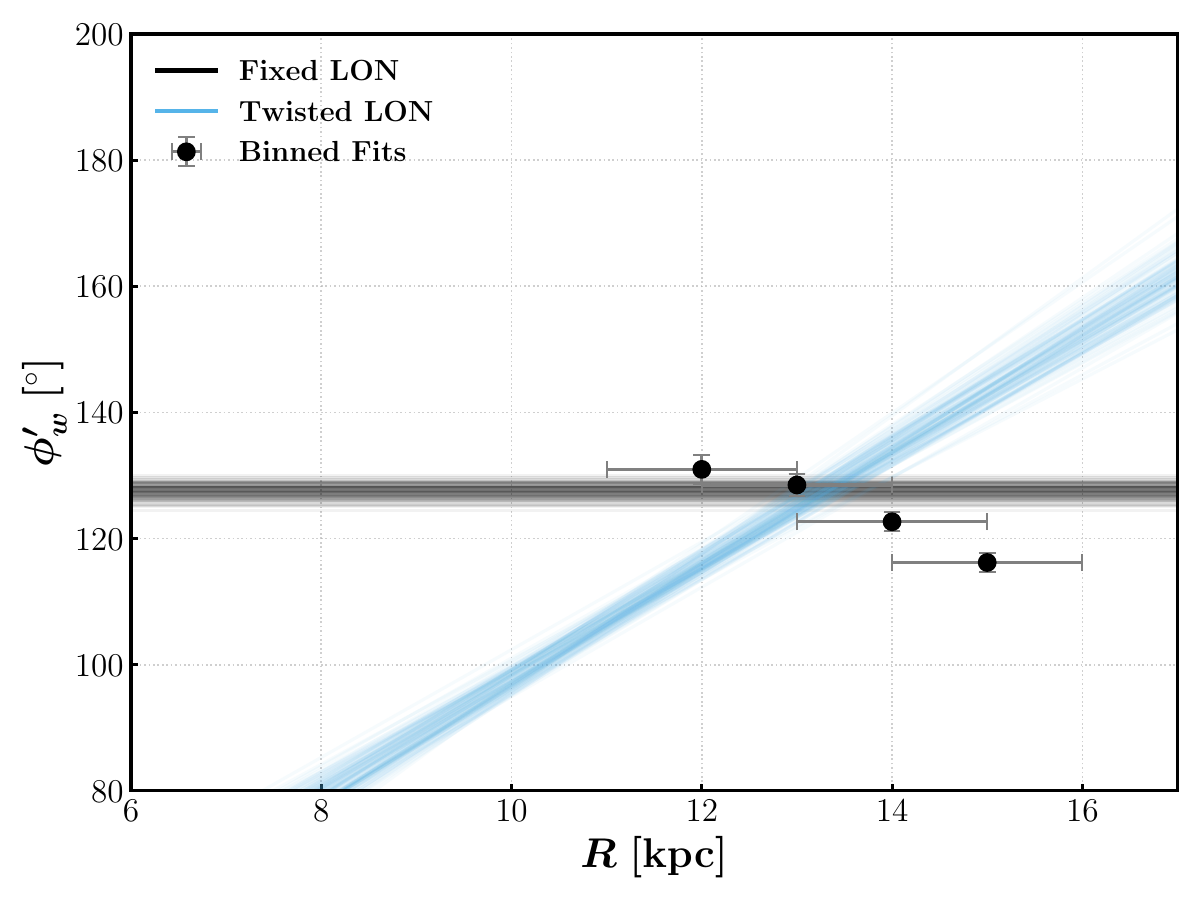}
    \caption{Galactic warp's line of nodes (LON) as a function of azimuthal angle ($\phi$) against Galactocentric radius ($R$) constructed from 100 random realisations of the MCMC posterior. The fixed model (black) exhibits a constant line of nodes across the entire disc, while the twisted model (blue) reveals a LON that resembles a trailing spiral. Black points represent independent fixed model fits to 2 kpc radial annuli from 11-13 kpc to 14-16 kpc. Vertical error bars denote 1$\sigma$ standard deviation of the posterior for each bin, horizontal error bars indicate the bin width.}
    \label{fig:LONplot} 
\end{figure}

\begin{figure}
	\includegraphics[width=\columnwidth]{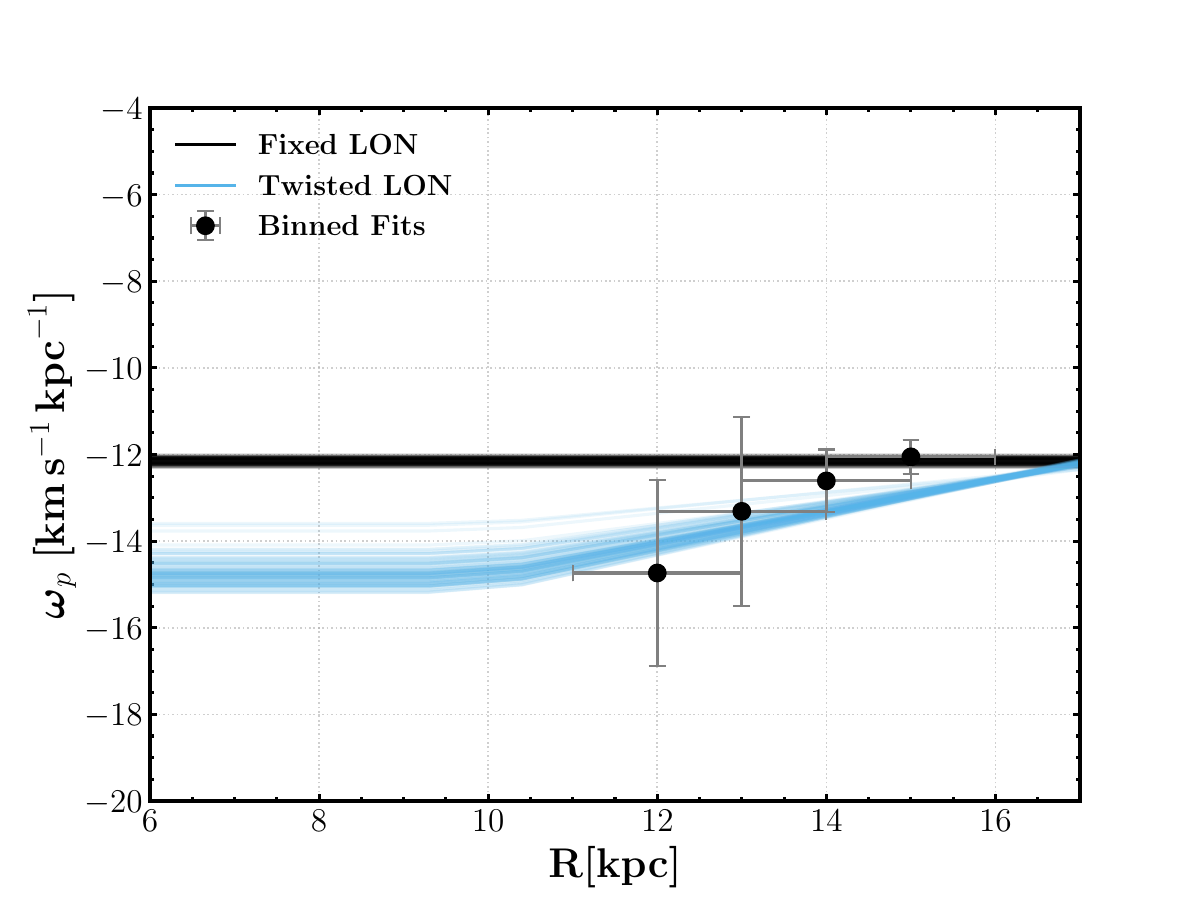}
    \caption{Precession rates of the fixed model (black line) and the twisted model (blue line) against Galactocentric radius. Both lines are constructed from 100 random realisations of the MCMC posterior distribution. The black points represent independent fixed model fits to 2 kpc radial annuli from 11-13 kpc to 14-16 kpc. Vertical error bars denote 1$\sigma$ standard deviation of the posterior for each bin, horizontal error bars indicate the bin width.}
    \label{fig:Omegaplot} 
\end{figure}

In Figures \ref{fig:LONplot} and \ref{fig:Omegaplot} we show the properties of the warp models. In each case we have also fit warp models in $2\kpc$ annuli, and shown these results as well. Figure \ref{fig:LONplot} shows the best-fitting parameters for the LON of each model. Our fixed model finds a LON of $128^\circ$, which is a prograde shift of $\sim10^\circ$ compared to the value found for the classical model in Paper I. There is a similar prograde offset in the twisted models. While both studies find a trailing spiral LON, they differ in the magnitude of the twist: our twisted model yields a gradient of $9^\circ\,\rm kpc^{-1}$, whereas Paper I reports a more moderate $5.2^\circ\,\rm kpc^{-1}$. 

However, as in Paper I, when our data is split into 2 kpc annuli, the resulting LON azimuth decreases with increasing Galactocentric radius, forming a leading spiral. This contrasts with the trailing spiral identified by the global twisted models. It is this continued discrepancy that reinforces the likelihood that the twisted model is too rigid and may force a trailing spiral to compensate for model shortcomings.

Figure \ref{fig:Omegaplot} displays the precession rates, $\omega_p(R)$, for the fixed and twisted model, with the same fixed-model analysis of 2 kpc annuli. Our fixed model's result of -12.2 km\,s$^{-1}\, \rm kpc^{-1}$ closely aligns with the $\sim$-13.3 km\,s$^{-1}\, \rm kpc^{-1}$ found in Paper I. However, the precession rate differs more from that found in Paper I when allowed to vary with Galactocentric radius. The twisted model in this study has a stronger differential precession, starting at -14.6 km\,s$^{-1}\, \rm kpc^{-1}$ at 10 kpc and increasing by -0.35 km\,s$^{-1}\, \rm kpc^{-2}$ in the retrograde direction. Conversely, Paper I conclude a precession rate that exhibits a negligible change with radius, maintaining a base rate of -13.4 km\,s$^{-1}\, \rm kpc^{-1}$ with a near-zero gradient of 0.02 km\,s$^{-1}\, \rm kpc^{-2}$. The 2 kpc split annuli have similar precession rates to those found at those radii for twisted model, despite having somewhat different LONs at these radii. 



\begin{figure}
	\includegraphics[width=\columnwidth]{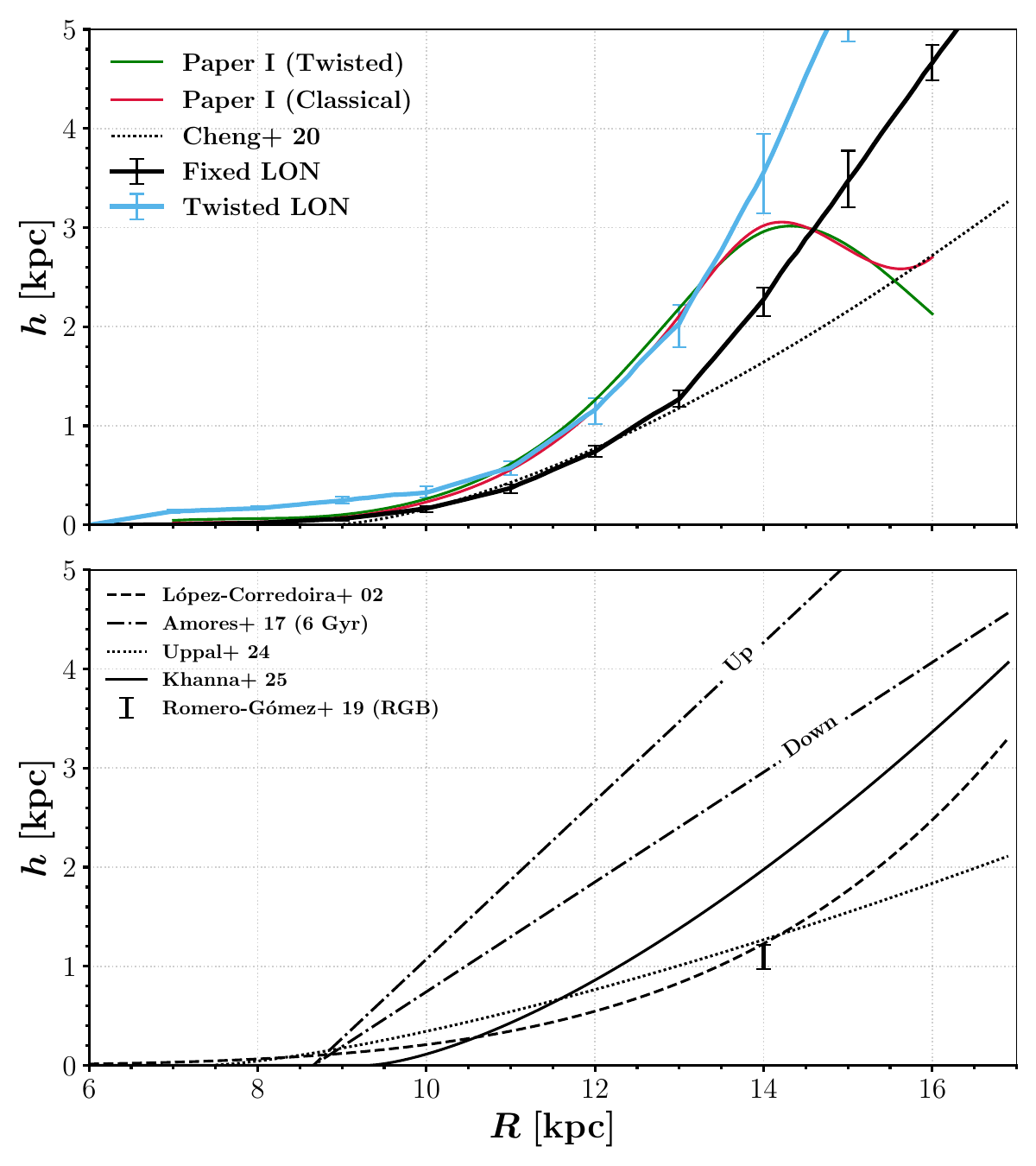}
    \caption{Top panel: Maximum warp amplitude as a function of Galactocentric radius for the fixed and twisted models. Solid lines represent the median values derived from the models' MCMC posteriors, with the error bars showing the $1\sigma$ confidence intervals. The twisted LON (blue line) and fixed LON (black line) both display fast growth of the warp in the outer disc reaching $5.56\pm0.43$ kpc and $4.71\pm0.22$ kpc at 16 kpc respectively. Bottom Panel: Comparative warp shapes reconstructed from various recent literature findings.}
    \label{fig:Heights} 
\end{figure}

The maximum warp amplitudes derived from our kinematic models show one major difference between our RC sample and the broad stellar sample utilised in Paper I. The upper panel in Fig.~\ref{fig:Heights} shows the warp shapes for the fixed and twisted models compared to models from Paper I. These models continue to ever larger amplitudes at larger radii, unlike the models in Paper I which plateaued beyond $14\kpc$. For further comparison we reproduce the warp amplitude from \cite{cheng_exploring_2020}, who also use kinematic data to fit their model, in the upper panel of Fig.~\ref{fig:Heights}; and warp amplitudes from recent studies that use star-count data of older populations in the lower panel \citep{lopez_corredoira_2mass_2002,amores_evolution_2017,romero-gomez_gaia_2019,uppal_clump_2024,khanna_unlimited_2025}.

Both our fixed LON and model and twisted LON model exhibit steep bending of the disc reaching 4.7 and 5.6 kpc at a radius of 16 kpc respectively, with the twisted model showing earlier bending surpassing the fixed model in the outer disc. In contrast, both classical models in Paper I reach a maximum warp amplitude of $\sim 3$ kpc near a radius of 14 kpc, before plateauing and slightly decreasing towards the outer disc. The steep warp of these models is most similar to the older stellar populations defined in \cite{amores_evolution_2017}, though they remain notably steeper than most other comparison models.

While both \citet{lopez_corredoira_2mass_2002} and \citet{khanna_unlimited_2025} find continuous warp growth in the outer regions, their warp amplitudes remain relatively small, reaching only $\sim 1.2$ kpc and $\sim 2.0$ kpc at $R=14$ kpc, respectively. \citet{cheng_exploring_2020} find a similar warp shape to that resulted from the fixed model up to a radius of 13 kpc, however does not continue to grow at the accelerated rate in the outer disc. With this, the amplitude measured by \citealt{romero-gomez_gaia_2019} using RGB stars remains lower than both of our global models.

Beyond the comparison models shown in Figure \ref{fig:Heights}, \citet{levine_vertical_2006} use H{\textsc i} mapping to find the warp reaching up to 6 kpc in the outer disc. \citet{reyle_2mass_2009} used 2MASS star counts to model the outer disc, finding a continuous, non-plateauing warp that mirrors our steep amplitude. Similarly, spatial mappings using M-giant stars \citep{momany_outer_2006} and K giants \citep{wang_nonsteady_2020} confirm that these older, dynamically relaxed tracers have sustained large warp amplitudes at large Galactocentric radii.

These results differ from those found using younger stellar populations. Mapping using Classical Cepheids \citep{chen_intuitive_2019, skowron_mapping_2019,lemasle_tracing_2022, cabrera_cepheids_2024} consistently show a shallower warp profile.

\section{Discussion} \label{sec:discussion}
In this study, we have traced the kinematic signature of the Milky Way's warp out to a Galactocentric radius beyond $16\kpc$ using a sample of RC stars. Using kinematic models holding both the LON and precession rate fixed across the disc or allowing them to vary linearly with Galactocentric radius, we have found that the red-clump warp is precessing rapidly and that its LON is offset from the generally accepted value based on the positions of these stars, near the Galactic anticentre \citep[e.g.,][K25]{uppal_clump_2024}. The precession of the disc is made particularly clear by the characteristic dip of average vertical velocity $\overline{v}_Z$ to negative values beyond $R\,{\sim}\,17\kpc$.

However, the differing results found under different modelling assumptions mean that we are still unable to assess with confidence the shape of the LON for the RC, and whether it is twisting in the same sense as disc rotation, as has been found for the warp traced by Cepheid variables \citep[e.g.,][]{chen_intuitive_2019, dehnen_a_twisted_2023, cabrera_cepheids_2024} and in external galaxies \citep{briggs_rules_1990}. Our twisted model twists in the opposite sense to disc rotation, while if we break the data into radial bins and fit each separately, their LONs twist in the same sense as the rotation. This is, frustratingly, similar to the results of Paper I and is probably a statement about the inflexibility of the models usually applied to warped discs. 


This study has taken the approach of Paper I and applied it to a carefully selected sample of RC stars from K25. It is therefore an improvement on Paper I in both data quality and homogeneity of the sample. 




Paper I analysed a broad, mixed stellar sample of roughly 27 million stars relying on photogeometric distances \citep{bailer-jones_estimating_2021} using \Gaia\ EDR3 data. In the Galactic midplane, where the line-of-sight is severely obscured by extinction, uncertainties on these distances based on photometry at optical wavelengths become substantial. By intentionally restricting our dataset to RC stars, and using distance estimates based on \AllWISE\ mid-infrared geometry, we have been able reduce distance uncertainties and mitigate the systematic uncertainties introduced by extinction.

As well as this improved distance mapping, restricting the sample to red clump stars means that we consider equivalent populations across the disc. Because the Paper I dataset spans the entire HR diagram, it inevitably favoured intrinsically brighter stars more the further away we looked from the Sun. This will have led to us sampling different age demographics in different parts of the disc. The RC sample we use in this paper contains stars with a wide range of ages, but does so in a consistent way across the disc. Since the behaviour of the warp appears to be a function of stellar age, this is a significant improvement.

\subsection{Model Limitations} \label{ssec:limitations}

\noindent While the models used in this paper have, for consistency with the wider literature, largely been adopted from Paper I, they have limitations and it is clear that some refinements would be necessary to increase physical realism. The simple sine-wave form for the warp is nearly equivalent to a tilted-ring model, but does not reflect the lop-sidedness of the Galactic warp \citep[e.g.,][]{amores_evolution_2017, romero-gomez_gaia_2019}, nor any radial corrugations \citep[e.g.,][]{xu_rings_2015}. 

The relatively limited spatial range that our data covers makes it hard to confidently disentangle these multiple terms, due to degeneracies. Experiments including a second-order Fourier term ($m=2$) did not produce satisfactory results, so we do not include these in this study.

One feature of the data that our models struggled to capture was the negative $\overline{v}_Z$ for stars at $\phi\gtrsim210^\circ$ even at $R\lesssim10\kpc$ (e.g., Fig.~\ref{fig:ModelDataComp}). This is a surprising feature of the data as it implies a substantial downward perturbation of the galaxy in this region at, and even within, the Sun's orbit around the Galactic centre. Similar perturbations are seen in two papers which accompanied \Gaia\ data releases \citep{Gaia_katz_18,gaia_drimmel_2023}, and in the latter case are dismissed as likely systematic errors. This is starting to seem less-and-less likely with further improvements to the data, including the improved distance estimates used in this paper. We emphasise that this is seen even when we consider only stars with measured line-of-sight velocities and (Fig.~\ref{fig:v_los_rangeincrease}, left panel) or if we restrict the sample to stars either above or below the plane.

It may be this feature of the data that explains the difference between the twist of the LON in cases where we allow it to vary linearly and cases where we fit to limited radial ranges in the disc. The model with a twisted LON reaches negative $\overline{v}_Z$ in this part of the galaxy (Fig.~\ref{fig:ModelDataComp}), and may be forcing the trailing line-of-nodes to fit this part of the data. This comes at the cost of, in places, a worse fit at larger radii.

One additional simplification that almost all modelling of the warp's kinematics makes is that the vertical velocity distribution of the stellar population is well captured by an average at a given $R$ and $\phi$. However, the \Gaia\ phase spiral \citep{antoja_dynamically_2018} means that the average $\overline{v}_Z$ is dependant on $Z$. We know that the phase spiral becomes particularly strong in parts of the outer disc \citep{alinder_investigating_2023,frankel_unwinding_2023} and this is likely to mean that simply looking at the average $v_Z$ is not fairly reflecting the velocity distribution. A modelling approach that unites the phase spiral with the Galactic warp would appear to be an important next step in our understanding of the disturbed Milky Way.

\subsection{Kinematics and Shape of the Warp} \label{ssec:kinematicsofthewarp}

\noindent Consistently, whatever tracers one uses, the Milky Way's disc appears warped in its outer radii. This warp is upwards in roughly the direction the Sun is currently moving and downwards on the opposite side of the Galaxy. The kinematics of the warp, in all tracers, indicate that it is precessing rapidly in the same direction as disc rotation \citep[e.g.,][Paper I]{poggio_evidence_2020, cheng_exploring_2020, dehnen_a_twisted_2023}.

Beyond these shared characteristics of all tracer populations, the story becomes harder to disentangle. For tracers of the young population, such as Cepheid variables, the amplitude of the warp is typically around $0.7\kpc$ at $R\,{\sim}\,14\kpc$ \citep[e.g.,][]{chen_intuitive_2019,skowron_three-dimensional_2019, poggio_greatwave_2025}, with a velocity signature that is nearly consistent with the positional one \citep{dehnen_a_twisted_2023} or offset slightly in the sense that the velocity signature trails the LON \citep{cabrera_cepheids_2024}. For older populations, such as the red clump stars, the warp is larger (see Fig~\ref{fig:Heights} for examples), and we have shown in this study that the LON from the velocity signature is offset by over $50^\circ$ from the LON found in studies of the positions of RC stars \citep[e.g.,][]{uppal_clump_2024}, in the sense that the velocity signature leads the LON.

The differences between populations of different ages, the rapid precession and the offsets in the LONs all appear to argue for a warp that is a response to a perturbation and is not likely to be long lived. The gas disc, from which the Cepheids relatively recently formed, will have been perturbed slightly differently to the existing stars, and this may explain the differences. If this is the case then we are left with the open question of why so many other galaxies appear to be warped. Are their warps longer lasting, or are a large fraction of them being perturbed enough to create a warp? And what is the key perturber for the Milky Way? Along with the more obvious candidates of the Sagittarius dwarf and Large Magellanic Cloud \citep[e.g.,][Astrophel \& McMillan in prep.]{laporte_response_2018,poggio_measuring_2021}, a misaligned dark matter halo might be a key consideration \citep{han_tilted_2023} though the velocity signature of this perturbation is not yet clear.

\section{Conclusions} \label{sec:conclusions}

\noindent This work looks at the kinematics of RC stars in the Galactic warp, and finds that they are consistent with a rapidly precessing warp, with a line of nodes that is offset by around $50^\circ$ from what we would expect from the stellar positions. The red clump stars are taken from the 
the catalogue of of K25, with distances taking advantage of \AllWISE\ photometry, combined with proper motions and line-of-sight velocities from from \Gaia\ DR3. 
We greatly increase the effective spatial range of the sample by estimating line-of-sight velocities of stars under the approximation of zero Galactocentric radial velocity for around 4 million stars, typically those that are fainter and further from the Sun. 

The RC warp is rapidly precessing, with a pattern speed of around 12 $\kms\kpc^{-1}$. If we allow it to vary with radius, then the pattern speed decreases slightly with radius from ${\sim}\,15\kms\kpc^{-1}$ at $R=10\kpc$ to ${\sim}\,12\kms\kpc^{-1}$ at $17\kpc$. This trend is in the same direction as that seen for Cepheids by \cite{dehnen_a_twisted_2023} and \cite{cabrera_cepheids_2024}, but the rate of change with radius is much smaller.

The amplitude of the warp continues to increase out to the greatest radii probed by our models, whether the LON is twisted or not. This is in contrast to the plateauing LON found by Paper I, which flattened out by $R\sim14\kpc$ at a maximum height of around $3\kpc$. The height reached in the outer radii is in excess of those found by studies of the positions of stars in the disc, including K25. It is reasonable to see this as another way in which the kinematics of the sample are inconsistent with the expectations from a simple model. The feature of the data that we can intuitively see requires this large amplitude warp is the steep decline in $\overline{v}_Z$ at large radii seen most clearly in Fig.~\ref{fig:ModelDataComp}.

Considering the velocity distribution by itself, as we have done in this study, is a very useful way of gaining insight into the system. It has the enormous benefit of allowing us to work in a way that is much less hampered by selection effects, and to determine what we would expect, given a kinematic model, for the distribution of stars. 
However, to gain physical insight we need to consider the spatial and kinematic distributions together, and find a model that can explain how they relate to one another. This is severely hampered by dust extinction, which prevents clear analysis of \Gaia\ data near the Galactic plane. Even careful modelling of the selection effects that apply to the \Gaia\ data did not allow a clear determination of the RC line-of-nodes from the positions of stars (K25). Future \Gaia\ data releases will suffer from the same effects, and it may require a future astrometric mission in the infrared \citep[e.g., {\it GaiaNIR}:][]{hobbs_all-sky_2021} to resolve this problem by seeing through the dust to the obscured portions of the Galaxy beyond.

\section*{Acknowledgements}

    The authors thank Shourya Khanna for providing early access to the red clump sample. This work has made use of data from the European Space Agency (ESA) mission
    {\it Gaia} (\url{https://www.cosmos.esa.int/gaia}), processed by the {\it Gaia}
    Data Processing and Analysis Consortium (DPAC,
    \url{https://www.cosmos.esa.int/web/gaia/dpac/consortium}). Funding for the DPAC
    has been provided by national institutions, in particular the institutions
    participating in the {\it Gaia} Multilateral Agreement. 
    This research has made use of NASA's Astrophysics Data System. A portion of this work used the ALICE high performance computing facility at the University of Leicester.
    This paper made use of the following software packages for Python:
    \textsc{AstroPy} \citep{astropy:2013,astropy:2018,astropy:2022}, 
    \textsc{emcee} \citep{foreman-mackey_emcee_2013}, 
    \textsc{Matplotlib} \citep{Hunter:2007}, 
	\textsc{Numpy} \citep{harris2020array},
    \textsc{Pandas} \citep{mckinney-proc-scipy-2010}, 
    and \textsc{SciPy} \citep{2020SciPy-NMeth}.
\section*{Data Availability}

Data used in this study was taken from \Gaia\ data release 3, available via \url{https://gea.esac.esa.int/archive/} and the red clump star catalogue from \cite{khanna_unlimited_2025} available at
the CDS via \url{https://cdsarc.cds.unistra.fr/viz-bin/cat/J/A+A/701/A270}. Code to produce the estimated velocities for stars without measured line-of-sight velocities is available upon reasonable request from the corresponding author.




\bibliographystyle{mnras}
\bibliography{references} 

@ARTICLE{antoja_dynamically_2018,
       author = {{Antoja}, T. and {Helmi}, A. and {Romero-G{\'o}mez}, M. and {Katz}, D. and {Babusiaux}, C. and {Drimmel}, R. and {Evans}, D.~W. and {Figueras}, F. and {Poggio}, E. and {Reyl{\'e}}, C. and {Robin}, A.~C. and {Seabroke}, G. and {Soubiran}, C.},
        title = "{A dynamically young and perturbed Milky Way disk}",
      journal = {\nat},
         year = 2018,
        month = sep,
       volume = {561},
       number = {7723},
        pages = {360-362},
          doi = {10.1038/s41586-018-0510-7},
archivePrefix = {arXiv},
       eprint = {1804.10196},
 primaryClass = {astro-ph.GA},
       adsurl = {https://ui.adsabs.harvard.edu/abs/2018Natur.561..360A}
}

@ARTICLE{alinder_investigating_2023,
       author = {{Alinder}, S. and {McMillan}, P.~J. and {Bensby}, T.},
        title = "{Investigating the amplitude and rotation of the phase spiral in the Milky Way outer disc}",
      journal = {\aap},
         year = 2023,
        month = oct,
       volume = {678},
          eid = {A46},
        pages = {A46},
          doi = {10.1051/0004-6361/202346560},
archivePrefix = {arXiv},
       eprint = {2303.18040},
 primaryClass = {astro-ph.GA},
       adsurl = {https://ui.adsabs.harvard.edu/abs/2023A&A...678A..46A}
}

@article{hobbs_all-sky_2021,
	title = {All-sky visible and near infrared space astrometry},
	volume = {51},
	issn = {0922-6435, 1572-9508},
	url = {https://link.springer.com/10.1007/s10686-021-09705-z},
	doi = {10.1007/s10686-021-09705-z},
	language = {en},
	number = {3},
	urldate = {2023-05-03},
	journal = {Experimental Astronomy},
	author = {Hobbs, David and Brown, Anthony and Høg, Erik and Jordi, Carme and Kawata, Daisuke and Tanga, Paolo and Klioner, Sergei and Sozzetti, Alessandro and Wyrzykowski, Łukasz and Walton, Nicholas and Vallenari, Antonella and Makarov, Valeri and Rybizki, Jan and Jiménez-Esteban, Fran and Caballero, José A. and McMillan, Paul J. and Secrest, Nathan and Mor, Roger and Andrews, Jeff J. and Zwitter, Tomaž and Chiappini, Cristina and Fynbo, Johan P. U. and Ting, Yuan-Sen and Hestroffer, Daniel and Lindegren, Lennart and McArthur, Barbara and Gouda, Naoteru and Moore, Anna and Gonzalez, Oscar A. and Vaccari, Mattia},
	month = jun,
	year = {2021},
	pages = {783--843},
}

@article{laporte_footprints_2019,
	title = {Footprints of the {Sagittarius} dwarf galaxy in the {Gaia} data set},
	volume = {485},
	issn = {0035-8711, 1365-2966},
	url = {https://academic.oup.com/mnras/article/485/3/3134/5366747},
	doi = {10.1093/mnras/stz583},
	language = {en},
	number = {3},
	urldate = {2023-04-18},
	journal = {\mnras},
	author = {Laporte, Chervin F P and Minchev, Ivan and Johnston, Kathryn V and Gómez, Facundo A},
	month = may,
	year = {2019},
	pages = {3134--3152},
}

@article{jeans_theory_1915,
	title = {On the {Theory} of {Star}-{Streaming} and the {Structure} of the {Universe}},
	volume = {76},
	issn = {0035-8711, 1365-2966},
	url = {https://academic.oup.com/mnras/article-lookup/doi/10.1093/mnras/76.2.70},
	doi = {10.1093/mnras/76.2.70},
	language = {en},
	number = {2},
	urldate = {2023-04-17},
	journal = {\mnras},
	author = {Jeans, J. H.},
	month = dec,
	year = {1915},
	pages = {70--84},
}

@article{briggs_rules_1990,
	title = {Rules of behavior for galactic {WARPS}},
	volume = {352},
	issn = {0004-637X, 1538-4357},
	url = {http://adsabs.harvard.edu/doi/10.1086/168512},
	doi = {10.1086/168512},
	language = {en},
	urldate = {2023-02-13},
	journal = {\apj},
	author = {Briggs, F. H.},
	month = mar,
	year = {1990},
	pages = {15},
}

@article{kerr_magellanic_1957,
	title = {A {Magellanic} effect on the galaxy},
	volume = {62},
	journal = {\aj},
	author = {Kerr, F. J.},
	month = may,
	year = {1957},
	pages = {93--93},
}

@article{gaia_collaboration_mission_2016,
       author = {{Gaia Collaboration} and {Prusti}, T. and {de Bruijne}, J.~H.~J. and {Brown}, A.~G.~A. and {Vallenari}, A. and {Babusiaux}, C. and {Bailer-Jones}, C.~A.~L. and {Bastian}, U. and {Biermann}, M. and {Evans}, D.~W. and {Eyer}, L. and {Jansen}, F. and {Jordi}, C. and {Klioner}, S.~A. and {Lammers}, U. and {Lindegren}, L. and {Luri}, X. and {Mignard}, F. and {Milligan}, D.~J. and {Panem}, C. and {Poinsignon}, V. and {Pourbaix}, D. and {Randich}, S. and {Sarri}, G. and {Sartoretti}, P. and {Siddiqui}, H.~I. and {Soubiran}, C. and {Valette}, V. and {van Leeuwen}, F. and {Walton}, N.~A. and {Aerts}, C. and {Arenou}, F. and {Cropper}, M. and {Drimmel}, R. and {H{\o}g}, E. and {Katz}, D. and {Lattanzi}, M.~G. and {O'Mullane}, W. and {Grebel}, E.~K. and {Holland}, A.~D. and {Huc}, C. and {Passot}, X. and {Bramante}, L. and {Cacciari}, C. and {Casta{\~n}eda}, J. and {Chaoul}, L. and {Cheek}, N. and {De Angeli}, F. and {Fabricius}, C. and {Guerra}, R. and {Hern{\'a}ndez}, J. and {Jean-Antoine-Piccolo}, A. and {Masana}, E. and {Messineo}, R. and {Mowlavi}, N. and {Nienartowicz}, K. and {Ord{\'o}{\~n}ez-Blanco}, D. and {Panuzzo}, P. and {Portell}, J. and {Richards}, P.~J. and {Riello}, M. and {Seabroke}, G.~M. and {Tanga}, P. and {Th{\'e}venin}, F. and {Torra}, J. and {Els}, S.~G. and {Gracia-Abril}, G. and {Comoretto}, G. and {Garcia-Reinaldos}, M. and {Lock}, T. and {Mercier}, E. and {Altmann}, M. and {Andrae}, R. and {Astraatmadja}, T.~L. and {Bellas-Velidis}, I. and {Benson}, K. and {Berthier}, J. and {Blomme}, R. and {Busso}, G. and {Carry}, B. and {Cellino}, A. and {Clementini}, G. and {Cowell}, S. and {Creevey}, O. and {Cuypers}, J. and {Davidson}, M. and {De Ridder}, J. and {de Torres}, A. and {Delchambre}, L. and {Dell'Oro}, A. and {Ducourant}, C. and {Fr{\'e}mat}, Y. and {Garc{\'\i}a-Torres}, M. and {Gosset}, E. and {Halbwachs}, J. -L. and {Hambly}, N.~C. and {Harrison}, D.~L. and {Hauser}, M. and {Hestroffer}, D. and {Hodgkin}, S.~T. and {Huckle}, H.~E. and {Hutton}, A. and {Jasniewicz}, G. and {Jordan}, S. and {Kontizas}, M. and {Korn}, A.~J. and {Lanzafame}, A.~C. and {Manteiga}, M. and {Moitinho}, A. and {Muinonen}, K. and {Osinde}, J. and {Pancino}, E. and {Pauwels}, T. and {Petit}, J. -M. and {Recio-Blanco}, A. and {Robin}, A.~C. and {Sarro}, L.~M. and {Siopis}, C. and {Smith}, M. and {Smith}, K.~W. and {Sozzetti}, A. and {Thuillot}, W. and {van Reeven}, W. and {Viala}, Y. and {Abbas}, U. and {Abreu Aramburu}, A. and {Accart}, S. and {Aguado}, J.~J. and {Allan}, P.~M. and {Allasia}, W. and {Altavilla}, G. and {{\'A}lvarez}, M.~A. and {Alves}, J. and {Anderson}, R.~I. and {Andrei}, A.~H. and {Anglada Varela}, E. and {Antiche}, E. and {Antoja}, T. and {Ant{\'o}n}, S. and {Arcay}, B. and {Atzei}, A. and {Ayache}, L. and {Bach}, N. and {Baker}, S.~G. and {Balaguer-N{\'u}{\~n}ez}, L. and {Barache}, C. and {Barata}, C. and {Barbier}, A. and {Barblan}, F. and {Baroni}, M. and {Barrado y Navascu{\'e}s}, D. and {Barros}, M. and {Barstow}, M.~A. and {Becciani}, U. and {Bellazzini}, M. and {Bellei}, G. and {Bello Garc{\'\i}a}, A. and {Belokurov}, V. and {Bendjoya}, P. and {Berihuete}, A. and {Bianchi}, L. and {Bienaym{\'e}}, O. and {Billebaud}, F. and {Blagorodnova}, N. and {Blanco-Cuaresma}, S. and {Boch}, T. and {Bombrun}, A. and {Borrachero}, R. and {Bouquillon}, S. and {Bourda}, G. and {Bouy}, H. and {Bragaglia}, A. and {Breddels}, M.~A. and {Brouillet}, N. and {Br{\"u}semeister}, T. and {Bucciarelli}, B. and {Budnik}, F. and {Burgess}, P. and {Burgon}, R. and {Burlacu}, A. and {Busonero}, D. and {Buzzi}, R. and {Caffau}, E. and {Cambras}, J. and {Campbell}, H. and {Cancelliere}, R. and {Cantat-Gaudin}, T. and {Carlucci}, T. and {Carrasco}, J.~M. and {Castellani}, M. and {Charlot}, P. and {Charnas}, J. and {Charvet}, P. and {Chassat}, F. and {Chiavassa}, A. and {Clotet}, M. and {Cocozza}, G. and {Collins}, R.~S. and {Collins}, P. and {Costigan}, G. and {Crifo}, F. and {Cross}, N.~J.~G. and {Crosta}, M. and {Crowley}, C. and {Dafonte}, C. and {Damerdji}, Y. and {Dapergolas}, A. and {David}, P. and {David}, M. and {De Cat}, P. and {de Felice}, F. and {de Laverny}, P. and {De Luise}, F. and {De March}, R. and {de Martino}, D. and {de Souza}, R. and {Debosscher}, J. and {del Pozo}, E. and {Delbo}, M. and {Delgado}, A. and {Delgado}, H.~E. and {di Marco}, F. and {Di Matteo}, P. and {Diakite}, S. and {Distefano}, E. and {Dolding}, C. and {Dos Anjos}, S. and {Drazinos}, P. and {Dur{\'a}n}, J. and {Dzigan}, Y. and {Ecale}, E. and {Edvardsson}, B. and {Enke}, H. and {Erdmann}, M. and {Escolar}, D. and {Espina}, M. and {Evans}, N.~W. and {Eynard Bontemps}, G. and {Fabre}, C. and {Fabrizio}, M. and {Faigler}, S. and {Falc{\~a}o}, A.~J. and {Farr{\`a}s Casas}, M. and {Faye}, F. and {Federici}, L. and {Fedorets}, G. and {Fern{\'a}ndez-Hern{\'a}ndez}, J. and {Fernique}, P. and {Fienga}, A. and {Figueras}, F. and {Filippi}, F. and {Findeisen}, K. and {Fonti}, A. and {Fouesneau}, M. and {Fraile}, E. and {Fraser}, M. and {Fuchs}, J. and {Furnell}, R. and {Gai}, M. and {Galleti}, S. and {Galluccio}, L. and {Garabato}, D. and {Garc{\'\i}a-Sedano}, F. and {Gar{\'e}}, P. and {Garofalo}, A. and {Garralda}, N. and {Gavras}, P. and {Gerssen}, J. and {Geyer}, R. and {Gilmore}, G. and {Girona}, S. and {Giuffrida}, G. and {Gomes}, M. and {Gonz{\'a}lez-Marcos}, A. and {Gonz{\'a}lez-N{\'u}{\~n}ez}, J. and {Gonz{\'a}lez-Vidal}, J.~J. and {Granvik}, M. and {Guerrier}, A. and {Guillout}, P. and {Guiraud}, J. and {G{\'u}rpide}, A. and {Guti{\'e}rrez-S{\'a}nchez}, R. and {Guy}, L.~P. and {Haigron}, R. and {Hatzidimitriou}, D. and {Haywood}, M. and {Heiter}, U. and {Helmi}, A. and {Hobbs}, D. and {Hofmann}, W. and {Holl}, B. and {Holland}, G. and {Hunt}, J.~A.~S. and {Hypki}, A. and {Icardi}, V. and {Irwin}, M. and {Jevardat de Fombelle}, G. and {Jofr{\'e}}, P. and {Jonker}, P.~G. and {Jorissen}, A. and {Julbe}, F. and {Karampelas}, A. and {Kochoska}, A. and {Kohley}, R. and {Kolenberg}, K. and {Kontizas}, E. and {Koposov}, S.~E. and {Kordopatis}, G. and {Koubsky}, P. and {Kowalczyk}, A. and {Krone-Martins}, A. and {Kudryashova}, M. and {Kull}, I. and {Bachchan}, R.~K. and {Lacoste-Seris}, F. and {Lanza}, A.~F. and {Lavigne}, J. -B. and {Le Poncin-Lafitte}, C. and {Lebreton}, Y. and {Lebzelter}, T. and {Leccia}, S. and {Leclerc}, N. and {Lecoeur-Taibi}, I. and {Lemaitre}, V. and {Lenhardt}, H. and {Leroux}, F. and {Liao}, S. and {Licata}, E. and {Lindstr{\o}m}, H.~E.~P. and {Lister}, T.~A. and {Livanou}, E. and {Lobel}, A. and {L{\"o}ffler}, W. and {L{\'o}pez}, M. and {Lopez-Lozano}, A. and {Lorenz}, D. and {Loureiro}, T. and {MacDonald}, I. and {Magalh{\~a}es Fernandes}, T. and {Managau}, S. and {Mann}, R.~G. and {Mantelet}, G. and {Marchal}, O. and {Marchant}, J.~M. and {Marconi}, M. and {Marie}, J. and {Marinoni}, S. and {Marrese}, P.~M. and {Marschalk{\'o}}, G. and {Marshall}, D.~J. and {Mart{\'\i}n-Fleitas}, J.~M. and {Martino}, M. and {Mary}, N. and {Matijevi{\v{c}}}, G. and {Mazeh}, T. and {McMillan}, P.~J. and {Messina}, S. and {Mestre}, A. and {Michalik}, D. and {Millar}, N.~R. and {Miranda}, B.~M.~H. and {Molina}, D. and {Molinaro}, R. and {Molinaro}, M. and {Moln{\'a}r}, L. and {Moniez}, M. and {Montegriffo}, P. and {Monteiro}, D. and {Mor}, R. and {Mora}, A. and {Morbidelli}, R. and {Morel}, T. and {Morgenthaler}, S. and {Morley}, T. and {Morris}, D. and {Mulone}, A.~F. and {Muraveva}, T. and {Musella}, I. and {Narbonne}, J. and {Nelemans}, G. and {Nicastro}, L. and {Noval}, L. and {Ord{\'e}novic}, C. and {Ordieres-Mer{\'e}}, J. and {Osborne}, P. and {Pagani}, C. and {Pagano}, I. and {Pailler}, F. and {Palacin}, H. and {Palaversa}, L. and {Parsons}, P. and {Paulsen}, T. and {Pecoraro}, M. and {Pedrosa}, R. and {Pentik{\"a}inen}, H. and {Pereira}, J. and {Pichon}, B. and {Piersimoni}, A.~M. and {Pineau}, F. -X. and {Plachy}, E. and {Plum}, G. and {Poujoulet}, E. and {Pr{\v{s}}a}, A. and {Pulone}, L. and {Ragaini}, S. and {Rago}, S. and {Rambaux}, N. and {Ramos-Lerate}, M. and {Ranalli}, P. and {Rauw}, G. and {Read}, A. and {Regibo}, S. and {Renk}, F. and {Reyl{\'e}}, C. and {Ribeiro}, R.~A. and {Rimoldini}, L. and {Ripepi}, V. and {Riva}, A. and {Rixon}, G. and {Roelens}, M. and {Romero-G{\'o}mez}, M. and {Rowell}, N. and {Royer}, F. and {Rudolph}, A. and {Ruiz-Dern}, L. and {Sadowski}, G. and {Sagrist{\`a} Sell{\'e}s}, T. and {Sahlmann}, J. and {Salgado}, J. and {Salguero}, E. and {Sarasso}, M. and {Savietto}, H. and {Schnorhk}, A. and {Schultheis}, M. and {Sciacca}, E. and {Segol}, M. and {Segovia}, J.~C. and {Segransan}, D. and {Serpell}, E. and {Shih}, I. -C. and {Smareglia}, R. and {Smart}, R.~L. and {Smith}, C. and {Solano}, E. and {Solitro}, F. and {Sordo}, R. and {Soria Nieto}, S. and {Souchay}, J. and {Spagna}, A. and {Spoto}, F. and {Stampa}, U. and {Steele}, I.~A. and {Steidelm{\"u}ller}, H. and {Stephenson}, C.~A. and {Stoev}, H. and {Suess}, F.~F. and {S{\"u}veges}, M. and {Surdej}, J. and {Szabados}, L. and {Szegedi-Elek}, E. and {Tapiador}, D. and {Taris}, F. and {Tauran}, G. and {Taylor}, M.~B. and {Teixeira}, R. and {Terrett}, D. and {Tingley}, B. and {Trager}, S.~C. and {Turon}, C. and {Ulla}, A. and {Utrilla}, E. and {Valentini}, G. and {van Elteren}, A. and {Van Hemelryck}, E. and {van Leeuwen}, M. and {Varadi}, M. and {Vecchiato}, A. and {Veljanoski}, J. and {Via}, T. and {Vicente}, D. and {Vogt}, S. and {Voss}, H. and {Votruba}, V. and {Voutsinas}, S. and {Walmsley}, G. and {Weiler}, M. and {Weingrill}, K. and {Werner}, D. and {Wevers}, T. and {Whitehead}, G. and {Wyrzykowski}, {\L}. and {Yoldas}, A. and {{\v{Z}}erjal}, M. and {Zucker}, S. and {Zurbach}, C. and {Zwitter}, T. and {Alecu}, A. and {Allen}, M. and {Allende Prieto}, C. and {Amorim}, A. and {Anglada-Escud{\'e}}, G. and {Arsenijevic}, V. and {Azaz}, S. and {Balm}, P. and {Beck}, M. and {Bernstein}, H. -H. and {Bigot}, L. and {Bijaoui}, A. and {Blasco}, C. and {Bonfigli}, M. and {Bono}, G. and {Boudreault}, S. and {Bressan}, A. and {Brown}, S. and {Brunet}, P. -M. and {Bunclark}, P. and {Buonanno}, R. and {Butkevich}, A.~G. and {Carret}, C. and {Carrion}, C. and {Chemin}, L. and {Ch{\'e}reau}, F. and {Corcione}, L. and {Darmigny}, E. and {de Boer}, K.~S. and {de Teodoro}, P. and {de Zeeuw}, P.~T. and {Delle Luche}, C. and {Domingues}, C.~D. and {Dubath}, P. and {Fodor}, F. and {Fr{\'e}zouls}, B. and {Fries}, A. and {Fustes}, D. and {Fyfe}, D. and {Gallardo}, E. and {Gallegos}, J. and {Gardiol}, D. and {Gebran}, M. and {Gomboc}, A. and {G{\'o}mez}, A. and {Grux}, E. and {Gueguen}, A. and {Heyrovsky}, A. and {Hoar}, J. and {Iannicola}, G. and {Isasi Parache}, Y. and {Janotto}, A. -M. and {Joliet}, E. and {Jonckheere}, A. and {Keil}, R. and {Kim}, D. -W. and {Klagyivik}, P. and {Klar}, J. and {Knude}, J. and {Kochukhov}, O. and {Kolka}, I. and {Kos}, J. and {Kutka}, A. and {Lainey}, V. and {LeBouquin}, D. and {Liu}, C. and {Loreggia}, D. and {Makarov}, V.~V. and {Marseille}, M.~G. and {Martayan}, C. and {Martinez-Rubi}, O. and {Massart}, B. and {Meynadier}, F. and {Mignot}, S. and {Munari}, U. and {Nguyen}, A. -T. and {Nordlander}, T. and {Ocvirk}, P. and {O'Flaherty}, K.~S. and {Olias Sanz}, A. and {Ortiz}, P. and {Osorio}, J. and {Oszkiewicz}, D. and {Ouzounis}, A. and {Palmer}, M. and {Park}, P. and {Pasquato}, E. and {Peltzer}, C. and {Peralta}, J. and {P{\'e}turaud}, F. and {Pieniluoma}, T. and {Pigozzi}, E. and {Poels}, J. and {Prat}, G. and {Prod'homme}, T. and {Raison}, F. and {Rebordao}, J.~M. and {Risquez}, D. and {Rocca-Volmerange}, B. and {Rosen}, S. and {Ruiz-Fuertes}, M.~I. and {Russo}, F. and {Sembay}, S. and {Serraller Vizcaino}, I. and {Short}, A. and {Siebert}, A. and {Silva}, H. and {Sinachopoulos}, D. and {Slezak}, E. and {Soffel}, M. and {Sosnowska}, D. and {Strai{\v{z}}ys}, V. and {ter Linden}, M. and {Terrell}, D. and {Theil}, S. and {Tiede}, C. and {Troisi}, L. and {Tsalmantza}, P. and {Tur}, D. and {Vaccari}, M. and {Vachier}, F. and {Valles}, P. and {Van Hamme}, W. and {Veltz}, L. and {Virtanen}, J. and {Wallut}, J. -M. and {Wichmann}, R. and {Wilkinson}, M.~I. and {Ziaeepour}, H. and {Zschocke}, S.},
        title = "{The Gaia mission}",
      journal = {\aap},
         year = 2016,
        month = nov,
       volume = {595},
          eid = {A1},
        pages = {A1},
          doi = {10.1051/0004-6361/201629272},
archivePrefix = {arXiv},
       eprint = {1609.04153},
 primaryClass = {astro-ph.IM},
       adsurl = {https://ui.adsabs.harvard.edu/abs/2016A&A...595A...1G}
}

@article{poggio_measuring_2021,
	title = {Measuring the vertical response of the {Galactic} disc to an infalling satellite},
	volume = {508},
	issn = {0035-8711, 1365-2966},
	url = {https://academic.oup.com/mnras/article/508/1/541/6343041},
	doi = {10.1093/mnras/stab2245},
	language = {en},
	number = {1},
	urldate = {2023-03-22},
	journal = {\mnras},
	author = {Poggio, Eloisa and Laporte, Chervin F P and Johnston, Kathryn V and D’Onghia, Elena and Drimmel, Ronald and Grion Filho, Douglas},
	month = sep,
	year = {2021},
	pages = {541--559},
}

@article{eilers_strength_2020,
	title = {The {Strength} of the {Dynamical} {Spiral} {Perturbation} in the {Galactic} {Disk}},
	volume = {900},
	issn = {1538-4357},
	url = {https://iopscience.iop.org/article/10.3847/1538-4357/abac0b},
	doi = {10.3847/1538-4357/abac0b},
	language = {en},
	number = {2},
	urldate = {2023-02-28},
	journal = {\apj},
	author = {Eilers, Anna-Christina and Hogg, David W. and Rix, Hans-Walter and Frankel, Neige and Hunt, Jason A. S. and Fouvry, Jean-Baptiste and Buck, Tobias},
	month = sep,
	year = {2020},
	pages = {186},
}

@article{poggio_evidence_2020,
	title = {Evidence of a dynamically evolving {Galactic} warp},
	volume = {4},
	issn = {2397-3366},
	url = {https://www.nature.com/articles/s41550-020-1017-3},
	doi = {10.1038/s41550-020-1017-3},
	language = {en},
	number = {6},
	urldate = {2023-02-27},
	journal = {Nature Astronomy},
	author = {Poggio, E. and Drimmel, R. and Andrae, R. and Bailer-Jones, C. A. L. and Fouesneau, M. and Lattanzi, M. G. and Smart, R. L. and Spagna, A.},
	month = mar,
	year = {2020},
	pages = {590--596},
}

@article{bailer-jones_estimating_2021,
       author = {{Bailer-Jones}, C.~A.~L. and {Rybizki}, J. and {Fouesneau}, M. and {Demleitner}, M. and {Andrae}, R.},
        title = "{Estimating Distances from Parallaxes. V. Geometric and Photogeometric Distances to 1.47 Billion Stars in Gaia Early Data Release 3}",
      journal = {\aj},
         year = 2021,
        month = mar,
       volume = {161},
       number = {3},
          eid = {147},
        pages = {147},
          doi = {10.3847/1538-3881/abd806},
archivePrefix = {arXiv},
       eprint = {2012.05220},
 primaryClass = {astro-ph.SR},
       adsurl = {https://ui.adsabs.harvard.edu/abs/2021AJ....161..147B}
}

@article{skowron_three-dimensional_2019,
	title = {A three-dimensional map of the {Milky} {Way} using classical {Cepheid} variable stars},
	volume = {365},
	issn = {0036-8075, 1095-9203},
	url = {http://arxiv.org/abs/1806.10653},
	doi = {10.1126/science.aau3181},
	number = {6452},
	urldate = {2023-02-13},
	journal = {Science},
	author = {Skowron, D. M. and Skowron, J. and Mróz, P. and Udalski, A. and Pietrukowicz, P. and Soszyński, I. and Szymański, M. K. and Poleski, R. and Kozłowski, S. and Ulaczyk, K. and Rybicki, K. and Iwanek, P.},
	month = aug,
	year = {2019},
	note = {arXiv:1806.10653 [astro-ph]},
	pages = {478--482},
}

@article{momany_outer_2006,
       author = {{Momany}, Y. and {Zaggia}, S. and {Gilmore}, G. and {Piotto}, G. and {Carraro}, G. and {Bedin}, L.~R. and {de Angeli}, F.},
        title = "{Outer structure of the Galactic warp and flare: explaining the Canis Major over-density}",
      journal = {\aap},
         year = 2006,
        month = may,
       volume = {451},
       number = {2},
        pages = {515-538},
          doi = {10.1051/0004-6361:20054081},
archivePrefix = {arXiv},
       eprint = {astro-ph/0603385},
 primaryClass = {astro-ph},
       adsurl = {https://ui.adsabs.harvard.edu/abs/2006A&A...451..515M}
}

@article{cheng_exploring_2020,
	title = {Exploring the {Galactic} {Warp} through {Asymmetries} in the {Kinematics} of the {Galactic} {Disk}},
	language = {en},
	journal = {\apj},
	author = {Cheng, Xinlun and Anguiano, Borja and Majewski, Steven R. and Hayws, Chrisitan and Arras, Phil and Chiappini, Cristina and Hasselquist, Sten and de Andrade Queiroz, Anna Barbara and Nitschelm, Christian and Garcia-Hernandez, Domingo Anibal and Lane, Richard R. and Roman-Lopes, Alexandre and Frinchaboy, Peter M.},
	year = {2020},
	pages = {15},
}

@article{drimmel_galactic_2000,
	title = {The {Galactic} warp in {OB} stars from {Hipparcos}},
	volume = {354},
	url = {http://arxiv.org/abs/astro-ph/9912398},
	language = {en},
	urldate = {2022-04-29},
	journal = {\aap},
	author = {Drimmel, R. and Smart, R. L. and Lattanzi, M. G.},
	month = feb,
	year = {2000},
	pages = {67--76},
}

@article{foreman-mackey_emcee_2013,
       author = {{Foreman-Mackey}, Daniel and {Hogg}, David W. and {Lang}, Dustin and {Goodman}, Jonathan},
        title = "{emcee: The MCMC Hammer}",
      journal = {\pasp},
         year = 2013,
        month = mar,
       volume = {125},
       number = {925},
        pages = {306},
          doi = {10.1086/670067},
archivePrefix = {arXiv},
       eprint = {1202.3665},
 primaryClass = {astro-ph.IM},
       adsurl = {https://ui.adsabs.harvard.edu/abs/2013PASP..125..306F}
}

@article{levine_vertical_2006,
	title = {The {Vertical} {Structure} of the {Outer} {Milky} {Way} {H} {\textless}span style="font-variant:small-caps;"{\textgreater}i{\textless}/span{\textgreater} {Disk}},
	volume = {643},
	issn = {0004-637X, 1538-4357},
	shorttitle = {The {Vertical} {Structure} of the {Outer} {Milky} {Way} {H} {\textless}span style="font-variant},
	url = {https://iopscience.iop.org/article/10.1086/503091},
	doi = {10.1086/503091},
	language = {en},
	number = {2},
	urldate = {2022-09-16},
	journal = {\apj},
	author = {Levine, E. S. and Blitz, Leo and Heiles, Carl},
	month = jun,
	year = {2006},
	pages = {881--896},
}

@ARTICLE{amores_evolution_2017,
       author = {{Am{\^o}res}, E.~B. and {Robin}, A.~C. and {Reyl{\'e}}, C.},
        title = "{Evolution over time of the Milky Way's disc shape}",
      journal = {\aap},
         year = 2017,
        month = jun,
       volume = {602},
          eid = {A67},
        pages = {A67},
          doi = {10.1051/0004-6361/201628461},
archivePrefix = {arXiv},
       eprint = {1701.00475},
 primaryClass = {astro-ph.GA},
       adsurl = {https://ui.adsabs.harvard.edu/abs/2017A&A...602A..67A}
}

@ARTICLE{chen_intuitive_2019,
       author = {{Chen}, Xiaodian and {Wang}, Shu and {Deng}, Licai and {de Grijs}, Richard and {Liu}, Chao and {Tian}, Hao},
        title = "{An intuitive 3D map of the Galactic warp's precession traced by classical Cepheids}",
      journal = {Nature Astronomy},
         year = 2019,
        month = feb,
       volume = {3},
        pages = {320-325},
          doi = {10.1038/s41550-018-0686-7},
archivePrefix = {arXiv},
       eprint = {1902.00998},
 primaryClass = {astro-ph.GA},
       adsurl = {https://ui.adsabs.harvard.edu/abs/2019NatAs...3..320C}
}

@article{skowron_mapping_2019,
	title = {Mapping the {Northern} {Galactic} {Disk} {Warp} with {Classical} {Cepheids}},
	volume = {69},
	issn = {00015237},
	url = {https://doi.org/10.32023/0001-5237/69.4.1},
	doi = {10.32023/0001-5237/69.4.1},
	language = {en},
	number = {4},
	urldate = {2022-04-29},
	journal = {Acta Astronomica},
	author = {Skowron, D.M. and Skowron, J. and Mróz, P. and Udalski, A. and Pietrukowicz, P. and Soszyński, I. and Szymański, M.K. and Poleski, R. and Kozłowski, S. and Ulaczyk, K. and Rybicki, K. and Iwanek, P. and Wrona, M. and Gromadzki, M.},
	month = dec,
	year = {2019},
	pages = {305--319},
}

@ARTICLE{gaia_collaboration_gaia_2021,
       author = {{Gaia Collaboration} and {Antoja}, T. and {McMillan}, P.~J. and {Kordopatis}, G. and {Ramos}, P. and {Helmi}, A. and {Balbinot}, E. and {Cantat-Gaudin}, T. and {Chemin}, L. and {Figueras}, F. and {Jordi}, C. and {Khanna}, S. and {Romero-G{\'o}mez}, M. and {Seabroke}, G.~M. and {Brown}, A.~G.~A. and {Vallenari}, A. and {Prusti}, T. and {de Bruijne}, J.~H.~J. and {Babusiaux}, C. and {Biermann}, M. and {Creevey}, O.~L. and {Evans}, D.~W. and {Eyer}, L. and {Hutton}, A. and {Jansen}, F. and {Klioner}, S.~A. and {Lammers}, U. and {Lindegren}, L. and {Luri}, X. and {Mignard}, F. and {Panem}, C. and {Pourbaix}, D. and {Randich}, S. and {Sartoretti}, P. and {Soubiran}, C. and {Walton}, N.~A. and {Arenou}, F. and {Bailer-Jones}, C.~A.~L. and {Bastian}, U. and {Cropper}, M. and {Drimmel}, R. and {Katz}, D. and {Lattanzi}, M.~G. and {van Leeuwen}, F. and {Bakker}, J. and {Casta{\~n}eda}, J. and {De Angeli}, F. and {Ducourant}, C. and {Fabricius}, C. and {Fouesneau}, M. and {Fr{\'e}mat}, Y. and {Guerra}, R. and {Guerrier}, A. and {Guiraud}, J. and {Jean-Antoine Piccolo}, A. and {Masana}, E. and {Messineo}, R. and {Mowlavi}, N. and {Nicolas}, C. and {Nienartowicz}, K. and {Pailler}, F. and {Panuzzo}, P. and {Riclet}, F. and {Roux}, W. and {Sordo}, R. and {Tanga}, P. and {Th{\'e}venin}, F. and {Gracia-Abril}, G. and {Portell}, J. and {Teyssier}, D. and {Altmann}, M. and {Andrae}, R. and {Bellas-Velidis}, I. and {Benson}, K. and {Berthier}, J. and {Blomme}, R. and {Brugaletta}, E. and {Burgess}, P.~W. and {Busso}, G. and {Carry}, B. and {Cellino}, A. and {Cheek}, N. and {Clementini}, G. and {Damerdji}, Y. and {Davidson}, M. and {Delchambre}, L. and {Dell'Oro}, A. and {Fern{\'a}ndez-Hern{\'a}ndez}, J. and {Galluccio}, L. and {Garc{\'\i}a-Lario}, P. and {Garcia-Reinaldos}, M. and {Gonz{\'a}lez-N{\'u}{\~n}ez}, J. and {Gosset}, E. and {Haigron}, R. and {Halbwachs}, J. -L. and {Hambly}, N.~C. and {Harrison}, D.~L. and {Hatzidimitriou}, D. and {Heiter}, U. and {Hern{\'a}ndez}, J. and {Hestroffer}, D. and {Hodgkin}, S.~T. and {Holl}, B. and {Jan{\ss}en}, K. and {Jevardat de Fombelle}, G. and {Jordan}, S. and {Krone-Martins}, A. and {Lanzafame}, A.~C. and {L{\"o}ffler}, W. and {Lorca}, A. and {Manteiga}, M. and {Marchal}, O. and {Marrese}, P.~M. and {Moitinho}, A. and {Mora}, A. and {Muinonen}, K. and {Osborne}, P. and {Pancino}, E. and {Pauwels}, T. and {Recio-Blanco}, A. and {Richards}, P.~J. and {Riello}, M. and {Rimoldini}, L. and {Robin}, A.~C. and {Roegiers}, T. and {Rybizki}, J. and {Sarro}, L.~M. and {Siopis}, C. and {Smith}, M. and {Sozzetti}, A. and {Ulla}, A. and {Utrilla}, E. and {van Leeuwen}, M. and {van Reeven}, W. and {Abbas}, U. and {Abreu Aramburu}, A. and {Accart}, S. and {Aerts}, C. and {Aguado}, J.~J. and {Ajaj}, M. and {Altavilla}, G. and {{\'A}lvarez}, M.~A. and {{\'A}lvarez Cid-Fuentes}, J. and {Alves}, J. and {Anderson}, R.~I. and {Varela}, E. Anglada and {Audard}, M. and {Baines}, D. and {Baker}, S.~G. and {Balaguer-N{\'u}{\~n}ez}, L. and {Balog}, Z. and {Barache}, C. and {Barbato}, D. and {Barros}, M. and {Barstow}, M.~A. and {Bartolom{\'e}}, S. and {Bassilana}, J. -L. and {Bauchet}, N. and {Baudesson-Stella}, A. and {Becciani}, U. and {Bellazzini}, M. and {Bernet}, M. and {Bertone}, S. and {Bianchi}, L. and {Blanco-Cuaresma}, S. and {Boch}, T. and {Bombrun}, A. and {Bossini}, D. and {Bouquillon}, S. and {Bragaglia}, A. and {Bramante}, L. and {Breedt}, E. and {Bressan}, A. and {Brouillet}, N. and {Bucciarelli}, B. and {Burlacu}, A. and {Busonero}, D. and {Butkevich}, A.~G. and {Buzzi}, R. and {Caffau}, E. and {Cancelliere}, R. and {C{\'a}novas}, H. and {Carballo}, R. and {Carlucci}, T. and {Carnerero}, M.~I. and {Carrasco}, J.~M. and {Casamiquela}, L. and {Castellani}, M. and {Castro-Ginard}, A. and {Castro Sampol}, P. and {Chaoul}, L. and {Charlot}, P. and {Chiavassa}, A. and {Cioni}, M. -R.~L. and {Comoretto}, G. and {Cooper}, W.~J. and {Cornez}, T. and {Cowell}, S. and {Crifo}, F. and {Crosta}, M. and {Crowley}, C. and {Dafonte}, C. and {Dapergolas}, A. and {David}, M. and {David}, P. and {de Laverny}, P. and {De Luise}, F. and {De March}, R. and {De Ridder}, J. and {de Souza}, R. and {de Teodoro}, P. and {de Torres}, A. and {del Peloso}, E.~F. and {del Pozo}, E. and {Delgado}, A. and {Delgado}, H.~E. and {Delisle}, J. -B. and {Di Matteo}, P. and {Diakite}, S. and {Diener}, C. and {Distefano}, E. and {Dolding}, C. and {Eappachen}, D. and {Enke}, H. and {Esquej}, P. and {Fabre}, C. and {Fabrizio}, M. and {Faigler}, S. and {Fedorets}, G. and {Fernique}, P. and {Fienga}, A. and {Fouron}, C. and {Fragkoudi}, F. and {Fraile}, E. and {Franke}, F. and {Gai}, M. and {Garabato}, D. and {Garcia-Gutierrez}, A. and {Garc{\'\i}a-Torres}, M. and {Garofalo}, A. and {Gavras}, P. and {Gerlach}, E. and {Geyer}, R. and {Giacobbe}, P. and {Gilmore}, G. and {Girona}, S. and {Giuffrida}, G. and {Gomez}, A. and {Gonzalez-Santamaria}, I. and {Gonz{\'a}lez-Vidal}, J.~J. and {Granvik}, M. and {Guti{\'e}rrez-S{\'a}nchez}, R. and {Guy}, L.~P. and {Hauser}, M. and {Haywood}, M. and {Hidalgo}, S.~L. and {Hilger}, T. and {H{\l}adczuk}, N. and {Hobbs}, D. and {Holland}, G. and {Huckle}, H.~E. and {Jasniewicz}, G. and {Jonker}, P.~G. and {Juaristi Campillo}, J. and {Julbe}, F. and {Karbevska}, L. and {Kervella}, P. and {Kochoska}, A. and {Kontizas}, M. and {Korn}, A.~J. and {Kostrzewa-Rutkowska}, Z. and {Kruszy{\'n}ska}, K. and {Lambert}, S. and {Lanza}, A.~F. and {Lasne}, Y. and {Le Campion}, J. -F. and {Le Fustec}, Y. and {Lebreton}, Y. and {Lebzelter}, T. and {Leccia}, S. and {Leclerc}, N. and {Lecoeur-Taibi}, I. and {Liao}, S. and {Licata}, E. and {Lindstr{\o}m}, H.~E.~P. and {Lister}, T.~A. and {Livanou}, E. and {Lobel}, A. and {Madrero Pardo}, P. and {Managau}, S. and {Mann}, R.~G. and {Marchant}, J.~M. and {Marconi}, M. and {Marcos Santos}, M.~M.~S. and {Marinoni}, S. and {Marocco}, F. and {Marshall}, D.~J. and {Martin Polo}, L. and {Mart{\'\i}n-Fleitas}, J.~M. and {Masip}, A. and {Massari}, D. and {Mastrobuono-Battisti}, A. and {Mazeh}, T. and {Messina}, S. and {Michalik}, D. and {Millar}, N.~R. and {Mints}, A. and {Molina}, D. and {Molinaro}, R. and {Moln{\'a}r}, L. and {Montegriffo}, P. and {Mor}, R. and {Morbidelli}, R. and {Morel}, T. and {Morris}, D. and {Mulone}, A.~F. and {Munoz}, D. and {Muraveva}, T. and {Murphy}, C.~P. and {Musella}, I. and {Noval}, L. and {Ord{\'e}novic}, C. and {Orr{\`u}}, G. and {Osinde}, J. and {Pagani}, C. and {Pagano}, I. and {Palaversa}, L. and {Palicio}, P.~A. and {Panahi}, A. and {Pawlak}, M. and {Pe{\~n}alosa Esteller}, X. and {Penttil{\"a}}, A. and {Piersimoni}, A.~M. and {Pineau}, F. -X. and {Plachy}, E. and {Plum}, G. and {Poggio}, E. and {Poretti}, E. and {Poujoulet}, E. and {Pr{\v{s}}a}, A. and {Pulone}, L. and {Racero}, E. and {Ragaini}, S. and {Rainer}, M. and {Raiteri}, C.~M. and {Rambaux}, N. and {Ramos-Lerate}, M. and {Re Fiorentin}, P. and {Regibo}, S. and {Reyl{\'e}}, C. and {Ripepi}, V. and {Riva}, A. and {Rixon}, G. and {Robichon}, N. and {Robin}, C. and {Roelens}, M. and {Rohrbasser}, L. and {Rowell}, N. and {Royer}, F. and {Rybicki}, K.~A. and {Sadowski}, G. and {Sagrist{\`a} Sell{\'e}s}, A. and {Sahlmann}, J. and {Salgado}, J. and {Salguero}, E. and {Samaras}, N. and {Sanchez Gimenez}, V. and {Sanna}, N. and {Santove{\~n}a}, R. and {Sarasso}, M. and {Schultheis}, M. and {Sciacca}, E. and {Segol}, M. and {Segovia}, J.~C. and {S{\'e}gransan}, D. and {Semeux}, D. and {Siddiqui}, H.~I. and {Siebert}, A. and {Siltala}, L. and {Slezak}, E. and {Smart}, R.~L. and {Solano}, E. and {Solitro}, F. and {Souami}, D. and {Souchay}, J. and {Spagna}, A. and {Spoto}, F. and {Steele}, I.~A. and {Steidelm{\"u}ller}, H. and {Stephenson}, C.~A. and {S{\"u}veges}, M. and {Szabados}, L. and {Szegedi-Elek}, E. and {Taris}, F. and {Tauran}, G. and {Taylor}, M.~B. and {Teixeira}, R. and {Thuillot}, W. and {Tonello}, N. and {Torra}, F. and {Torra}, J. and {Turon}, C. and {Unger}, N. and {Vaillant}, M. and {van Dillen}, E. and {Vanel}, O. and {Vecchiato}, A. and {Viala}, Y. and {Vicente}, D. and {Voutsinas}, S. and {Weiler}, M. and {Wevers}, T. and {Wyrzykowski}, {\L}. and {Yoldas}, A. and {Yvard}, P. and {Zhao}, H. and {Zorec}, J. and {Zucker}, S. and {Zurbach}, C. and {Zwitter}, T.},
        title = "{Gaia Early Data Release 3. The Galactic anticentre}",
      journal = {\aap},
         year = 2021,
        month = may,
       volume = {649},
          eid = {A8},
        pages = {A8},
          doi = {10.1051/0004-6361/202039714},
archivePrefix = {arXiv},
       eprint = {2101.05811},
 primaryClass = {astro-ph.GA},
       adsurl = {https://ui.adsabs.harvard.edu/abs/2021A&A...649A...8G}
}

@ARTICLE{gaia_collaboration_summary_2023,
       author = {{Gaia Collaboration} and {Vallenari}, A. and {Brown}, A.~G.~A. and {Prusti}, T. and {de Bruijne}, J.~H.~J. and {Arenou}, F. and {Babusiaux}, C. and {Biermann}, M. and {Creevey}, O.~L. and {Ducourant}, C. and {Evans}, D.~W. and {Eyer}, L. and {Guerra}, R. and {Hutton}, A. and {Jordi}, C. and {Klioner}, S.~A. and {Lammers}, U.~L. and {Lindegren}, L. and {Luri}, X. and {Mignard}, F. and {Panem}, C. and {Pourbaix}, D. and {Randich}, S. and {Sartoretti}, P. and {Soubiran}, C. and {Tanga}, P. and {Walton}, N.~A. and {Bailer-Jones}, C.~A.~L. and {Bastian}, U. and {Drimmel}, R. and {Jansen}, F. and {Katz}, D. and {Lattanzi}, M.~G. and {van Leeuwen}, F. and {Bakker}, J. and {Cacciari}, C. and {Casta{\~n}eda}, J. and {De Angeli}, F. and {Fabricius}, C. and {Fouesneau}, M. and {Fr{\'e}mat}, Y. and {Galluccio}, L. and {Guerrier}, A. and {Heiter}, U. and {Masana}, E. and {Messineo}, R. and {Mowlavi}, N. and {Nicolas}, C. and {Nienartowicz}, K. and {Pailler}, F. and {Panuzzo}, P. and {Riclet}, F. and {Roux}, W. and {Seabroke}, G.~M. and {Sordo}, R. and {Th{\'e}venin}, F. and {Gracia-Abril}, G. and {Portell}, J. and {Teyssier}, D. and {Altmann}, M. and {Andrae}, R. and {Audard}, M. and {Bellas-Velidis}, I. and {Benson}, K. and {Berthier}, J. and {Blomme}, R. and {Burgess}, P.~W. and {Busonero}, D. and {Busso}, G. and {C{\'a}novas}, H. and {Carry}, B. and {Cellino}, A. and {Cheek}, N. and {Clementini}, G. and {Damerdji}, Y. and {Davidson}, M. and {de Teodoro}, P. and {Nu{\~n}ez Campos}, M. and {Delchambre}, L. and {Dell'Oro}, A. and {Esquej}, P. and {Fern{\'a}ndez-Hern{\'a}ndez}, J. and {Fraile}, E. and {Garabato}, D. and {Garc{\'\i}a-Lario}, P. and {Gosset}, E. and {Haigron}, R. and {Halbwachs}, J. -L. and {Hambly}, N.~C. and {Harrison}, D.~L. and {Hern{\'a}ndez}, J. and {Hestroffer}, D. and {Hodgkin}, S.~T. and {Holl}, B. and {Jan{\ss}en}, K. and {Jevardat de Fombelle}, G. and {Jordan}, S. and {Krone-Martins}, A. and {Lanzafame}, A.~C. and {L{\"o}ffler}, W. and {Marchal}, O. and {Marrese}, P.~M. and {Moitinho}, A. and {Muinonen}, K. and {Osborne}, P. and {Pancino}, E. and {Pauwels}, T. and {Recio-Blanco}, A. and {Reyl{\'e}}, C. and {Riello}, M. and {Rimoldini}, L. and {Roegiers}, T. and {Rybizki}, J. and {Sarro}, L.~M. and {Siopis}, C. and {Smith}, M. and {Sozzetti}, A. and {Utrilla}, E. and {van Leeuwen}, M. and {Abbas}, U. and {{\'A}brah{\'a}m}, P. and {Abreu Aramburu}, A. and {Aerts}, C. and {Aguado}, J.~J. and {Ajaj}, M. and {Aldea-Montero}, F. and {Altavilla}, G. and {{\'A}lvarez}, M.~A. and {Alves}, J. and {Anders}, F. and {Anderson}, R.~I. and {Anglada Varela}, E. and {Antoja}, T. and {Baines}, D. and {Baker}, S.~G. and {Balaguer-N{\'u}{\~n}ez}, L. and {Balbinot}, E. and {Balog}, Z. and {Barache}, C. and {Barbato}, D. and {Barros}, M. and {Barstow}, M.~A. and {Bartolom{\'e}}, S. and {Bassilana}, J. -L. and {Bauchet}, N. and {Becciani}, U. and {Bellazzini}, M. and {Berihuete}, A. and {Bernet}, M. and {Bertone}, S. and {Bianchi}, L. and {Binnenfeld}, A. and {Blanco-Cuaresma}, S. and {Blazere}, A. and {Boch}, T. and {Bombrun}, A. and {Bossini}, D. and {Bouquillon}, S. and {Bragaglia}, A. and {Bramante}, L. and {Breedt}, E. and {Bressan}, A. and {Brouillet}, N. and {Brugaletta}, E. and {Bucciarelli}, B. and {Burlacu}, A. and {Butkevich}, A.~G. and {Buzzi}, R. and {Caffau}, E. and {Cancelliere}, R. and {Cantat-Gaudin}, T. and {Carballo}, R. and {Carlucci}, T. and {Carnerero}, M.~I. and {Carrasco}, J.~M. and {Casamiquela}, L. and {Castellani}, M. and {Castro-Ginard}, A. and {Chaoul}, L. and {Charlot}, P. and {Chemin}, L. and {Chiaramida}, V. and {Chiavassa}, A. and {Chornay}, N. and {Comoretto}, G. and {Contursi}, G. and {Cooper}, W.~J. and {Cornez}, T. and {Cowell}, S. and {Crifo}, F. and {Cropper}, M. and {Crosta}, M. and {Crowley}, C. and {Dafonte}, C. and {Dapergolas}, A. and {David}, M. and {David}, P. and {de Laverny}, P. and {De Luise}, F. and {De March}, R. and {De Ridder}, J. and {de Souza}, R. and {de Torres}, A. and {del Peloso}, E.~F. and {del Pozo}, E. and {Delbo}, M. and {Delgado}, A. and {Delisle}, J. -B. and {Demouchy}, C. and {Dharmawardena}, T.~E. and {Di Matteo}, P. and {Diakite}, S. and {Diener}, C. and {Distefano}, E. and {Dolding}, C. and {Edvardsson}, B. and {Enke}, H. and {Fabre}, C. and {Fabrizio}, M. and {Faigler}, S. and {Fedorets}, G. and {Fernique}, P. and {Fienga}, A. and {Figueras}, F. and {Fournier}, Y. and {Fouron}, C. and {Fragkoudi}, F. and {Gai}, M. and {Garcia-Gutierrez}, A. and {Garcia-Reinaldos}, M. and {Garc{\'\i}a-Torres}, M. and {Garofalo}, A. and {Gavel}, A. and {Gavras}, P. and {Gerlach}, E. and {Geyer}, R. and {Giacobbe}, P. and {Gilmore}, G. and {Girona}, S. and {Giuffrida}, G. and {Gomel}, R. and {Gomez}, A. and {Gonz{\'a}lez-N{\'u}{\~n}ez}, J. and {Gonz{\'a}lez-Santamar{\'\i}a}, I. and {Gonz{\'a}lez-Vidal}, J.~J. and {Granvik}, M. and {Guillout}, P. and {Guiraud}, J. and {Guti{\'e}rrez-S{\'a}nchez}, R. and {Guy}, L.~P. and {Hatzidimitriou}, D. and {Hauser}, M. and {Haywood}, M. and {Helmer}, A. and {Helmi}, A. and {Sarmiento}, M.~H. and {Hidalgo}, S.~L. and {Hilger}, T. and {H{\l}adczuk}, N. and {Hobbs}, D. and {Holland}, G. and {Huckle}, H.~E. and {Jardine}, K. and {Jasniewicz}, G. and {Jean-Antoine Piccolo}, A. and {Jim{\'e}nez-Arranz}, {\'O}. and {Jorissen}, A. and {Juaristi Campillo}, J. and {Julbe}, F. and {Karbevska}, L. and {Kervella}, P. and {Khanna}, S. and {Kontizas}, M. and {Kordopatis}, G. and {Korn}, A.~J. and {K{\'o}sp{\'a}l}, {\'A}. and {Kostrzewa-Rutkowska}, Z. and {Kruszy{\'n}ska}, K. and {Kun}, M. and {Laizeau}, P. and {Lambert}, S. and {Lanza}, A.~F. and {Lasne}, Y. and {Le Campion}, J. -F. and {Lebreton}, Y. and {Lebzelter}, T. and {Leccia}, S. and {Leclerc}, N. and {Lecoeur-Taibi}, I. and {Liao}, S. and {Licata}, E.~L. and {Lindstr{\o}m}, H.~E.~P. and {Lister}, T.~A. and {Livanou}, E. and {Lobel}, A. and {Lorca}, A. and {Loup}, C. and {Madrero Pardo}, P. and {Magdaleno Romeo}, A. and {Managau}, S. and {Mann}, R.~G. and {Manteiga}, M. and {Marchant}, J.~M. and {Marconi}, M. and {Marcos}, J. and {Marcos Santos}, M.~M.~S. and {Mar{\'\i}n Pina}, D. and {Marinoni}, S. and {Marocco}, F. and {Marshall}, D.~J. and {Martin Polo}, L. and {Mart{\'\i}n-Fleitas}, J.~M. and {Marton}, G. and {Mary}, N. and {Masip}, A. and {Massari}, D. and {Mastrobuono-Battisti}, A. and {Mazeh}, T. and {McMillan}, P.~J. and {Messina}, S. and {Michalik}, D. and {Millar}, N.~R. and {Mints}, A. and {Molina}, D. and {Molinaro}, R. and {Moln{\'a}r}, L. and {Monari}, G. and {Mongui{\'o}}, M. and {Montegriffo}, P. and {Montero}, A. and {Mor}, R. and {Mora}, A. and {Morbidelli}, R. and {Morel}, T. and {Morris}, D. and {Muraveva}, T. and {Murphy}, C.~P. and {Musella}, I. and {Nagy}, Z. and {Noval}, L. and {Oca{\~n}a}, F. and {Ogden}, A. and {Ordenovic}, C. and {Osinde}, J.~O. and {Pagani}, C. and {Pagano}, I. and {Palaversa}, L. and {Palicio}, P.~A. and {Pallas-Quintela}, L. and {Panahi}, A. and {Payne-Wardenaar}, S. and {Pe{\~n}alosa Esteller}, X. and {Penttil{\"a}}, A. and {Pichon}, B. and {Piersimoni}, A.~M. and {Pineau}, F. -X. and {Plachy}, E. and {Plum}, G. and {Poggio}, E. and {Pr{\v{s}}a}, A. and {Pulone}, L. and {Racero}, E. and {Ragaini}, S. and {Rainer}, M. and {Raiteri}, C.~M. and {Rambaux}, N. and {Ramos}, P. and {Ramos-Lerate}, M. and {Re Fiorentin}, P. and {Regibo}, S. and {Richards}, P.~J. and {Rios Diaz}, C. and {Ripepi}, V. and {Riva}, A. and {Rix}, H. -W. and {Rixon}, G. and {Robichon}, N. and {Robin}, A.~C. and {Robin}, C. and {Roelens}, M. and {Rogues}, H.~R.~O. and {Rohrbasser}, L. and {Romero-G{\'o}mez}, M. and {Rowell}, N. and {Royer}, F. and {Ruz Mieres}, D. and {Rybicki}, K.~A. and {Sadowski}, G. and {S{\'a}ez N{\'u}{\~n}ez}, A. and {Sagrist{\`a} Sell{\'e}s}, A. and {Sahlmann}, J. and {Salguero}, E. and {Samaras}, N. and {Sanchez Gimenez}, V. and {Sanna}, N. and {Santove{\~n}a}, R. and {Sarasso}, M. and {Schultheis}, M. and {Sciacca}, E. and {Segol}, M. and {Segovia}, J.~C. and {S{\'e}gransan}, D. and {Semeux}, D. and {Shahaf}, S. and {Siddiqui}, H.~I. and {Siebert}, A. and {Siltala}, L. and {Silvelo}, A. and {Slezak}, E. and {Slezak}, I. and {Smart}, R.~L. and {Snaith}, O.~N. and {Solano}, E. and {Solitro}, F. and {Souami}, D. and {Souchay}, J. and {Spagna}, A. and {Spina}, L. and {Spoto}, F. and {Steele}, I.~A. and {Steidelm{\"u}ller}, H. and {Stephenson}, C.~A. and {S{\"u}veges}, M. and {Surdej}, J. and {Szabados}, L. and {Szegedi-Elek}, E. and {Taris}, F. and {Taylor}, M.~B. and {Teixeira}, R. and {Tolomei}, L. and {Tonello}, N. and {Torra}, F. and {Torra}, J. and {Torralba Elipe}, G. and {Trabucchi}, M. and {Tsounis}, A.~T. and {Turon}, C. and {Ulla}, A. and {Unger}, N. and {Vaillant}, M.~V. and {van Dillen}, E. and {van Reeven}, W. and {Vanel}, O. and {Vecchiato}, A. and {Viala}, Y. and {Vicente}, D. and {Voutsinas}, S. and {Weiler}, M. and {Wevers}, T. and {Wyrzykowski}, {\L}. and {Yoldas}, A. and {Yvard}, P. and {Zhao}, H. and {Zorec}, J. and {Zucker}, S. and {Zwitter}, T.},
        title = "{Gaia Data Release 3. Summary of the content and survey properties}",
      journal = {\aap},
         year = 2023,
        month = jun,
       volume = {674},
          eid = {A1},
        pages = {A1},
          doi = {10.1051/0004-6361/202243940},
archivePrefix = {arXiv},
       eprint = {2208.00211},
 primaryClass = {astro-ph.GA},
       adsurl = {https://ui.adsabs.harvard.edu/abs/2023A&A...674A...1G}
}

@ARTICLE{katz_gaia_2023,
       author = {{Katz}, D. and {Sartoretti}, P. and {Guerrier}, A. and {Panuzzo}, P. and {Seabroke}, G.~M. and {Th{\'e}venin}, F. and {Cropper}, M. and {Benson}, K. and {Blomme}, R. and {Haigron}, R. and {Marchal}, O. and {Smith}, M. and {Baker}, S. and {Chemin}, L. and {Damerdji}, Y. and {David}, M. and {Dolding}, C. and {Fr{\'e}mat}, Y. and {Gosset}, E. and {Jan{\ss}en}, K. and {Jasniewicz}, G. and {Lobel}, A. and {Plum}, G. and {Samaras}, N. and {Snaith}, O. and {Soubiran}, C. and {Vanel}, O. and {Zwitter}, T. and {Antoja}, T. and {Arenou}, F. and {Babusiaux}, C. and {Brouillet}, N. and {Caffau}, E. and {Di Matteo}, P. and {Fabre}, C. and {Fabricius}, C. and {Fragkoudi}, F. and {Haywood}, M. and {Huckle}, H.~E. and {Hottier}, C. and {Lasne}, Y. and {Leclerc}, N. and {Mastrobuono-Battisti}, A. and {Royer}, F. and {Teyssier}, D. and {Zorec}, J. and {Crifo}, F. and {Jean-Antoine Piccolo}, A. and {Turon}, C. and {Viala}, Y.},
        title = "{Gaia Data Release 3. Properties and validation of the radial velocities}",
      journal = {\aap},
         year = 2023,
        month = jun,
       volume = {674},
          eid = {A5},
        pages = {A5},
          doi = {10.1051/0004-6361/202244220},
archivePrefix = {arXiv},
       eprint = {2206.05902},
 primaryClass = {astro-ph.GA},
       adsurl = {https://ui.adsabs.harvard.edu/abs/2023A&A...674A...5K}
}

@ARTICLE{lemasle_tracing_2022,
       author = {{Lemasle}, B. and {Lala}, H.~N. and {Kovtyukh}, V. and {Hanke}, M. and {Prudil}, Z. and {Bono}, G. and {Braga}, V.~F. and {da Silva}, R. and {Fabrizio}, M. and {Fiorentino}, G. and {Fran{\c{c}}ois}, P. and {Grebel}, E.~K. and {Kniazev}, A.},
        title = "{Tracing the Milky Way warp and spiral arms with classical Cepheids}",
      journal = {\aap},
         year = 2022,
        month = dec,
       volume = {668},
          eid = {A40},
        pages = {A40},
          doi = {10.1051/0004-6361/202243273},
archivePrefix = {arXiv},
       eprint = {2209.02731},
 primaryClass = {astro-ph.GA},
       adsurl = {https://ui.adsabs.harvard.edu/abs/2022A&A...668A..40L}
}

@ARTICLE{mcmillan_disturbed_2022,
       author = {{McMillan}, Paul J. and {Petersson}, Jonathan and {Tepper-Garcia}, Thor and {Bland-Hawthorn}, Joss and {Antoja}, Teresa and {Chemin}, Laurent and {Figueras}, Francesca and {Khanna}, Shourya and {Kordopatis}, Georges and {Ramos}, Pau and {Romero-G{\'o}mez}, Merce and {Seabroke}, George},
        title = "{The disturbed outer Milky Way disc}",
      journal = {\mnras},
         year = 2022,
        month = nov,
       volume = {516},
       number = {4},
        pages = {4988-5002},
          doi = {10.1093/mnras/stac2571},
archivePrefix = {arXiv},
       eprint = {2206.04059},
 primaryClass = {astro-ph.GA},
       adsurl = {https://ui.adsabs.harvard.edu/abs/2022MNRAS.516.4988M}
}

@article{romero-gomez_gaia_2019,
       author = {{Romero-G{\'o}mez}, M. and {Mateu}, C. and {Aguilar}, L. and {Figueras}, F. and {Castro-Ginard}, A.},
        title = "{Gaia kinematics reveal a complex lopsided and twisted Galactic disc warp}",
      journal = {\aap},
         year = 2019,
        month = jul,
       volume = {627},
          eid = {A150},
        pages = {A150},
          doi = {10.1051/0004-6361/201834908},
archivePrefix = {arXiv},
       eprint = {1812.07576},
 primaryClass = {astro-ph.GA},
       adsurl = {https://ui.adsabs.harvard.edu/abs/2019A&A...627A.150R}
}

@article{poggio_galactic_2018,
	title = {The {Galactic} warp revealed by {Gaia} {DR2} kinematics},
	volume = {481},
	issn = {1745-3925, 1745-3933},
	url = {https://academic.oup.com/mnrasl/article/481/1/L21/5071966},
	doi = {10.1093/mnrasl/sly148},
	language = {en},
	number = {1},
	urldate = {2022-04-29},
	journal = {\mnras: Letters},
	author = {Poggio, E and Drimmel, R and Lattanzi, M G and Smart, R L and Spagna, A and Andrae, R and Bailer-Jones, C A L and Fouesneau, M and Antoja, T and Babusiaux, C and Evans, D W and Figueras, F and Katz, D and Reylé, C and Robin, A C and Romero-Gómez, M and Seabroke, G M},
	month = nov,
	year = {2018},
	pages = {L21--L25},
}

@ARTICLE{dehnen_a_twisted_2023,
       author = {{Dehnen}, Walter and {Semczuk}, Marcin and {Sch{\"o}nrich}, Ralph},
        title = "{A twisted and precessing Cepheid warp in the outer Milky Way disc}",
      journal = {\mnras},
         year = 2023,
        month = jul,
       volume = {523},
       number = {1},
        pages = {1556-1564},
          doi = {10.1093/mnras/stad1502},
archivePrefix = {arXiv},
       eprint = {2305.09343},
 primaryClass = {astro-ph.GA},
       adsurl = {https://ui.adsabs.harvard.edu/abs/2023MNRAS.523.1556D}
}

@InProceedings{mckinney-proc-scipy-2010,
  author    = { {McKinney}, Wes },
  title     = { {D}ata {S}tructures for {S}tatistical {C}omputing in {P}ython },
  booktitle = { {P}roceedings of the 9th {P}ython in {S}cience {C}onference },
  pages     = { 56 - 61 },
  year      = { 2010 },
  editor    = { {S}t\'efan van der {W}alt and {J}arrod {M}illman },
  doi       = { 10.25080/Majora-92bf1922-00a }
}

@ARTICLE{goodman_ensample,
       author = {{Goodman}, Jonathan and {Weare}, Jonathan},
        title = "{Ensemble samplers with affine invariance}",
      journal = {Communications in Applied Mathematics and Computational Science},
         year = 2010,
        month = jan,
       volume = {5},
       number = {1},
        pages = {65-80},
          doi = {10.2140/camcos.2010.5.65},
       adsurl = {https://ui.adsabs.harvard.edu/abs/2010CAMCS...5...65G}
}

@ARTICLE{lopez_corredoira_2mass_2002,
       author = {{L{\'o}pez-Corredoira}, M. and {Cabrera-Lavers}, A. and {Garz{\'o}n}, F. and {Hammersley}, P.~L.},
        title = "{Old stellar Galactic disc in near-plane regions according to 2MASS: Scales, cut-off, flare and warp}",
      journal = {\aap},
         year = 2002,
        month = nov,
       volume = {394},
        pages = {883-899},
          doi = {10.1051/0004-6361:20021175},
archivePrefix = {arXiv},
       eprint = {astro-ph/0208236},
 primaryClass = {astro-ph},
       adsurl = {https://ui.adsabs.harvard.edu/abs/2002A&A...394..883L}
}

@ARTICLE{2018A&A...615L..15G,
       author = {{GRAVITY Collaboration} and {Abuter}, R. and {Amorim}, A. and {Anugu}, N. and {Baub{\"o}ck}, M. and {Benisty}, M. and {Berger}, J.~P. and {Blind}, N. and {Bonnet}, H. and {Brandner}, W. and {Buron}, A. and {Collin}, C. and {Chapron}, F. and {Cl{\'e}net}, Y. and {Coud{\'e} Du Foresto}, V. and {de Zeeuw}, P.~T. and {Deen}, C. and {Delplancke-Str{\"o}bele}, F. and {Dembet}, R. and {Dexter}, J. and {Duvert}, G. and {Eckart}, A. and {Eisenhauer}, F. and {Finger}, G. and {F{\"o}rster Schreiber}, N.~M. and {F{\'e}dou}, P. and {Garcia}, P. and {Garcia Lopez}, R. and {Gao}, F. and {Gendron}, E. and {Genzel}, R. and {Gillessen}, S. and {Gordo}, P. and {Habibi}, M. and {Haubois}, X. and {Haug}, M. and {Hau{\ss}mann}, F. and {Henning}, Th. and {Hippler}, S. and {Horrobin}, M. and {Hubert}, Z. and {Hubin}, N. and {Jimenez Rosales}, A. and {Jochum}, L. and {Jocou}, K. and {Kaufer}, A. and {Kellner}, S. and {Kendrew}, S. and {Kervella}, P. and {Kok}, Y. and {Kulas}, M. and {Lacour}, S. and {Lapeyr{\`e}re}, V. and {Lazareff}, B. and {Le Bouquin}, J. -B. and {L{\'e}na}, P. and {Lippa}, M. and {Lenzen}, R. and {M{\'e}rand}, A. and {M{\"u}ler}, E. and {Neumann}, U. and {Ott}, T. and {Palanca}, L. and {Paumard}, T. and {Pasquini}, L. and {Perraut}, K. and {Perrin}, G. and {Pfuhl}, O. and {Plewa}, P.~M. and {Rabien}, S. and {Ram{\'\i}rez}, A. and {Ramos}, J. and {Rau}, C. and {Rodr{\'\i}guez-Coira}, G. and {Rohloff}, R. -R. and {Rousset}, G. and {Sanchez-Bermudez}, J. and {Scheithauer}, S. and {Sch{\"o}ller}, M. and {Schuler}, N. and {Spyromilio}, J. and {Straub}, O. and {Straubmeier}, C. and {Sturm}, E. and {Tacconi}, L.~J. and {Tristram}, K.~R.~W. and {Vincent}, F. and {von Fellenberg}, S. and {Wank}, I. and {Waisberg}, I. and {Widmann}, F. and {Wieprecht}, E. and {Wiest}, M. and {Wiezorrek}, E. and {Woillez}, J. and {Yazici}, S. and {Ziegler}, D. and {Zins}, G.},
        title = "{Detection of the gravitational redshift in the orbit of the star S2 near the Galactic centre massive black hole}",
      journal = {\aap},
         year = 2018,
        month = jul,
       volume = {615},
          eid = {L15},
        pages = {L15},
          doi = {10.1051/0004-6361/201833718},
archivePrefix = {arXiv},
       eprint = {1807.09409},
 primaryClass = {astro-ph.GA},
       adsurl = {https://ui.adsabs.harvard.edu/abs/2018A&A...615L..15G}
}

@ARTICLE{williams_wobbly_2013,
       author = {{Williams}, M.~E.~K. and {Steinmetz}, M. and {Binney}, J. and {Siebert}, A. and {Enke}, H. and {Famaey}, B. and {Minchev}, I. and {de Jong}, R.~S. and {Boeche}, C. and {Freeman}, K.~C. and {Bienaym{\'e}}, O. and {Bland-Hawthorn}, J. and {Gibson}, B.~K. and {Gilmore}, G.~F. and {Grebel}, E.~K. and {Helmi}, A. and {Kordopatis}, G. and {Munari}, U. and {Navarro}, J.~F. and {Parker}, Q.~A. and {Reid}, W. and {Seabroke}, G.~M. and {Sharma}, S. and {Siviero}, A. and {Watson}, F.~G. and {Wyse}, R.~F.~G. and {Zwitter}, T.},
        title = "{The wobbly Galaxy: kinematics north and south with RAVE red-clump giants}",
      journal = {\mnras},
         year = 2013,
        month = nov,
       volume = {436},
       number = {1},
        pages = {101-121},
          doi = {10.1093/mnras/stt1522},
archivePrefix = {arXiv},
       eprint = {1302.2468},
 primaryClass = {astro-ph.GA},
       adsurl = {https://ui.adsabs.harvard.edu/abs/2013MNRAS.436..101W}
}

@ARTICLE{frankel_unwinding_2023,
       author = {{Frankel}, Neige and {Bovy}, Jo and {Tremaine}, Scott and {Hogg}, David W.},
        title = "{Vertical motion in the Galactic disc: unwinding the snail}",
      journal = {\mnras},
         year = 2023,
        month = jun,
       volume = {521},
       number = {4},
        pages = {5917-5926},
          doi = {10.1093/mnras/stad908},
archivePrefix = {arXiv},
       eprint = {2212.11991},
 primaryClass = {astro-ph.GA},
       adsurl = {https://ui.adsabs.harvard.edu/abs/2023MNRAS.521.5917F}
}

@Article{Hunter:2007,
  Author    = {Hunter, J. D.},
  Title     = {Matplotlib: A 2D graphics environment},
  Journal   = {Computing in Science \& Engineering},
  Volume    = {9},
  Number    = {3},
  Pages     = {90--95},
  publisher = {IEEE COMPUTER SOC},
  doi       = {10.1109/MCSE.2007.55},
  year      = 2007
}

@ARTICLE{Girardi_RC_2016,
       author = {{Girardi}, L{\'e}o},
        title = "{Red Clump Stars}",
      journal = {\araa},
         year = 2016,
        month = sep,
       volume = {54},
        pages = {95-133},
          doi = {10.1146/annurev-astro-081915-023354},
       adsurl = {https://ui.adsabs.harvard.edu/abs/2016ARA&A..54...95G}
}




\appendix

\section{Tests and checks}\label{sec:checks}

\noindent To greatly increase the effective spatial range of the dataset and improve model robustness, line-of-sight velocities were estimated under the approximation of zero Galactocentric radial velocity ($\overline{v}_{R,gal}=0$), as detailed in section \ref{sec:MissLOS}. To quantify the uncertainties introduced into the data from the use of this method, the same assumption was made on a random selection of $2\,000\,000$  sources from the mock dataset GeDR3Mock \citep{rybizki_mockdata_2020}. 
Since we have absolute magnitudes for the mock stars, we isolate the RC stars with a cut in  absolute magnitude  ($-0.16<M_G<1.04,-1.98<M_{K_S}<-1.38$), resulting in a total mock sample of 1 318 926 RC stars.

To test the robustness of our assumptions, we conducted two mock tests on the GeDR3Mock sample, perturbed in realistic ways, to consider the effect of on our estimates, similar to those conducted by \cite{mcmillan_disturbed_2022}.  

For one test, we perturbed the radial velocities of our mock dataset using an $m=2$ spiral perturbation described by:
\begin{equation}
    \Delta v_R(R,\phi) = v_{\rm max} \sin\left[m\left(\phi - \phi_0 - \frac{\log(R/\kpc)}{\tan(p)}\right)\right],
\end{equation}
\noindent where $v_{\rm max}=7\kms$, $\phi_0=0^\circ$, and pitch angle $p=12^\circ$. These parameters create a spiral perturbation comparable to that found in the outer disc by \cite{eilers_strength_2020}, or the $v_R$ variation mapped by \cite{cheng_exploring_2020}. 

For the second test, we perturb the sample to reflect the complicated vertical kinematic disequilibrium that has been throughout the Milky Way disc. This notably includes the \Gaia\ phase space spiral \citep{antoja_dynamically_2018}, which is a particularly strong perturbation of the disc's distribution in the outer Galaxy \citep{alinder_investigating_2023}. 
To simulate the vertical asymmetry, the vertical velocities of stars in the mock catalogue were were modified with a step function defined as
\begin{equation}
     \Delta v_Z(Z) = 
     \begin{cases} 
       +2 \text{ km s}^{-1} & \text{for } Z > 0 \\
       -2 \text{ km s}^{-1} & \text{for } Z < 0 
     \end{cases}
 \end{equation}
This split mimics the vertical velocity differences observed between stellar populations on either side of the Galactic midplane found by \cite{gaia_collaboration_gaia_2021}. This tests whether asymmetrical extinction in the Galactic disc biases the value of $\overline{v}_Z$ found at a given $R,\phi$ by a significant amount.

\begin{figure*}
	\includegraphics[width=\columnwidth]{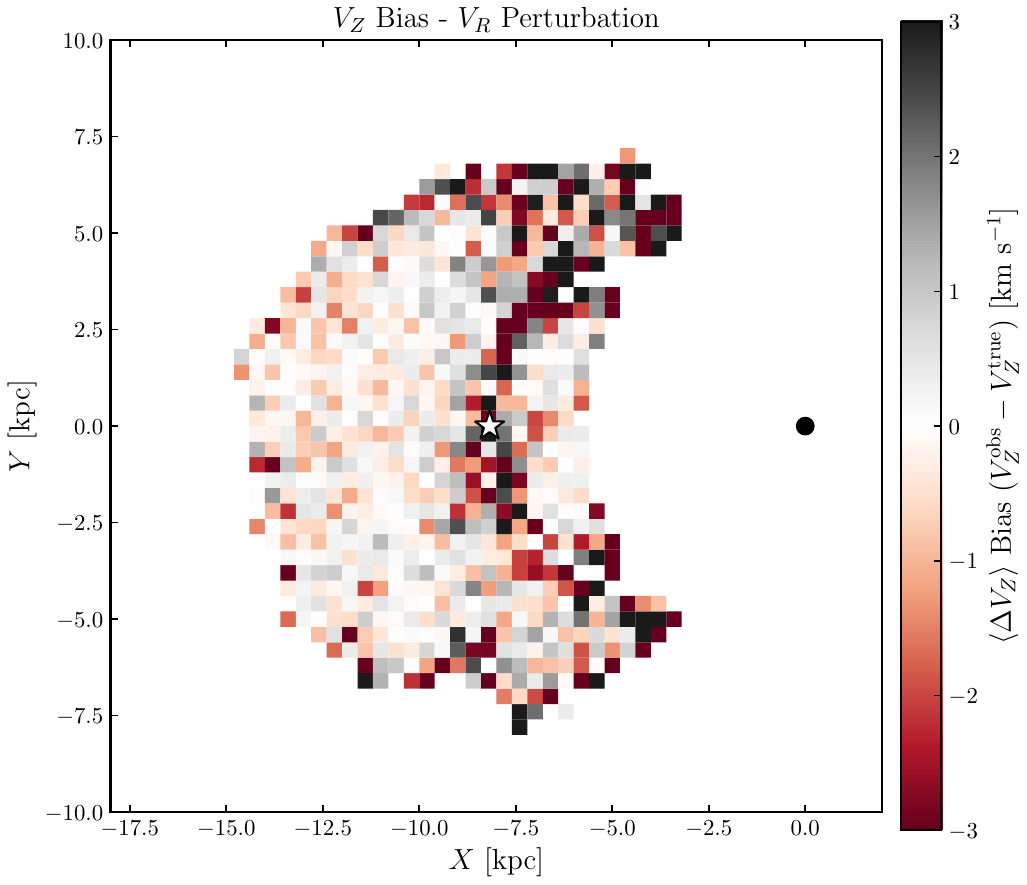}
    \includegraphics[width=\columnwidth]{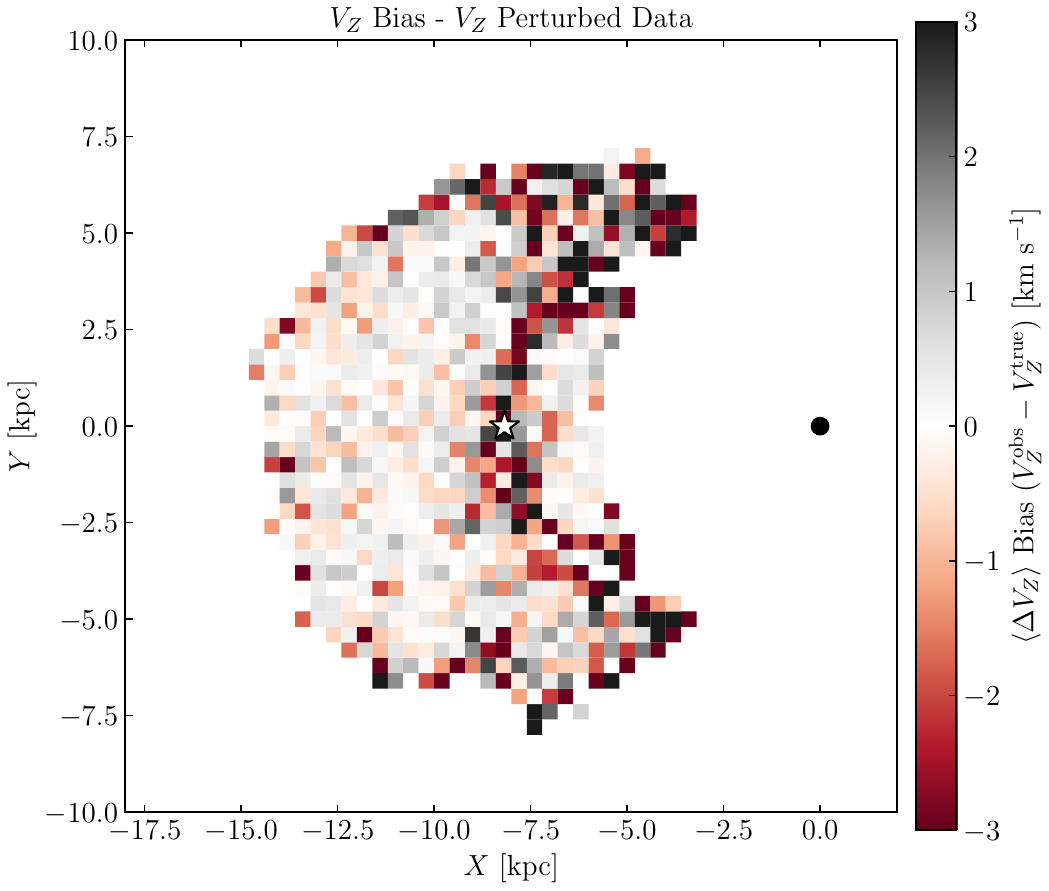}
    \caption{Spatial distribution of median $v_Z$ on the mock dataset perturbed with the $m=2$ spiral perturbation (left) or shift in vertical velocity at $Z=0$ (right). The propagated vertical velocity errors remain small in both cases.}
    \label{fig:ErrorsMock1} 
\end{figure*}

We applied our approximation of $\overline{v}_{R,gal}=0$ to the two mock datasets, and show the resulting $\overline{v}_Z$ biasses in Figure \ref{fig:ErrorsMock1}. These errors are generally small, however, notable exceptions occur near $l=90^\circ$ and $l=270^\circ$. These localised spikes are expected artifacts, occurring where the line-of-sight vector becomes nearly perpendicular to the Galactocentric radial velocity vector. The error in the region where we find our warp signature is small so we are comfortable assuming that the approximation is good enough for this work.


\section{Best Fitting Parameters}

\noindent Table \ref{bestfitting} provides the best-fitting values and associated uncertainties from the MCMC posterior distributions for the parameters of both the fixed and twisted kinematic models. Note that the LoN parameters have not been added to $180^\circ$ to show the symmetric warp closer to the Sun and are the direct results of the modelling.

\begin{table}
    \centering
    \caption{Best-fitting parameters for both fixed and twisted models with respective uncertainties.}
    \label{bestfitting}
    \begin{tabular}{lcc}
        \hline
        Parameter & Fixed model & Twisted model \\
        \hline
        $h_{7}$ [kpc] & $0.005 \pm 0.004$ & $0.140 \pm 0.016$ \\
        $h_{8}$ [kpc] & $0.020 \pm 0.011$ & $0.170 \pm 0.020$ \\
        $h_{9}$ [kpc] & $0.063 \pm 0.021$ & $0.252 \pm 0.033$ \\
        $h_{10}$ [kpc] & $0.162 \pm 0.031$ & $0.329 \pm 0.044$ \\
        $h_{11}$ [kpc] & $0.373 \pm 0.044$ & $0.578 \pm 0.080$ \\
        $h_{12}$ [kpc] & $0.739 \pm 0.058$ & $1.165 \pm 0.122$ \\
        $h_{13}$ [kpc] & $1.274 \pm 0.084$ & $2.039 \pm 0.188$ \\
        $h_{14}$ [kpc] & $2.272 \pm 0.137$ & $3.61 \pm 0.374$ \\
        $h_{15}$ [kpc] & $3.548 \pm 0.284$ & $5.389 \pm 0.428$ \\
        $h_{16}$ [kpc] & $4.710 \pm 0.218$ & $5.555 \pm 0.433$ \\
        \hline
        $\phi_{w}^{'}$ [deg] & $127.6 \pm 1.1$ & --- \\
        $\phi_{w,0}^{'}$ [deg] & --- & $97.9 \pm 1.4$ \\
        $\phi_{w,1}^{'}$ [deg kpc$^{-1}$] & --- & $9.0 \pm 0.5$ \\
        \hline
        $\omega_{p}$ [km s$^{-1}$ kpc$^{-1}$] & $-12.15 \pm 0.06$ & --- \\
        $\omega_{p,0}$ [km s$^{-1}$ kpc$^{-1}$] & --- & $-14.6 \pm 0.4$ \\
        $\omega_{p,1}$ [km s$^{-1}$ kpc$^{-2}$] & --- & $-0.35 \pm 0.06$ \\
        \hline
    \end{tabular}
\end{table}


\bsp	
\label{lastpage}
\end{document}